\RequirePackage{fix-cm}
\documentclass[twocolumn]{svjour3}          % twocolumn
\smartqed  % flush right qed marks, e.g. at end of proof
\usepackage{graphicx}
\usepackage{booktabs}
\usepackage{lipsum}
\usepackage{algorithm}
\usepackage{algpseudocode}
\usepackage{mathtools}
\usepackage{url}
\usepackage{amsmath,amsfonts, amssymb}
\usepackage{textcomp}
\usepackage{xcolor}
\usepackage[dvipsnames]{xcolor}
\usepackage{multirow}
\usepackage{enumitem}
\usepackage{tabularx}
\usepackage{adjustbox}
\usepackage{comment}
\usepackage{booktabs} % For professional table lines (\toprule, \midrule, \bottomrule)
\usepackage{array}
\usepackage{soul}
\usepackage{float}
\usepackage{fontawesome5}
\usepackage{makecell}
\usepackage{microtype}
\usepackage{hyperref}
\usepackage{ragged2e}
\hypersetup{hidelinks}
\hypersetup{
    colorlinks=true,
    linkcolor=black,
    citecolor=blue,
    urlcolor=blue
}

\newcommand{\orcid}[1]{\href{https://orcid.org/#1}{\textcolor[HTML]{A6CE39}{\faOrcid}}}

\begin{document}

\title{SoK: Secure Software-Based Multi-Domain Data Segregation%\thanks{Grants or other notes
%about the article that should go on the front page should be
%placed here. General acknowledgments should be placed at the end of the article.}
}
%\subtitle{Do you have a subtitle?\\ If so, write it here}

%\titlerunning{Short form of title}        % if too long for running head

\author{Quang Cao\textsuperscript{1} \and Peter Vinci\textsuperscript{2} \and Nick Georghiou\textsuperscript{2} \and Shane Phillips\textsuperscript{2} \and \\ Mathieu Philippe\textsuperscript{2} \and Nalin Arachchilage\textsuperscript{1} %etc.
}

%\authorrunning{Short form of author list} % if too long for running head

\institute{{\faEnvelope} Quang Cao \orcid{0000-0001-9649-943X} \at
              nhat.quang.cao2@rmit.edu.au          %  \\
%             \emph{Present address:} of F. Author  %  if needed
           \and
           %Peter Vinci \at
              %peter.vinci@c4i.com
           %\and
           %Nick Georghiou %\at
              %nick.georghiou@c4i.com
           %\and
           %Shane Phillips %\at
              %shane.phillips@c4i.com
           %\and
           %Mathieu Philippe %\at
              %mathieu.philippe@c4i.com
          %\and
           Nalin Arachchilage \orcid{0000-0002-0059-0376} \at
              nalin.arachchilage@rmit.edu.au
        \and \at
        \textsuperscript{1} RMIT University, Melbourne, Australia
        \and
        \textsuperscript{2} C4i Pty Ltd, Melbourne, Australia
}

\date{Received: date / Accepted: date}
% The correct dates will be entered by the editor

\maketitle

\begin{abstract}
Modern mission-critical coordination demands seamless communication across multiple domains. Traditionally, Voice Communication Systems (VCS) have relied on physically separated Red/Black architectures to ensure voice and data segregation. While these hardware-based methods provide strong security assurances and remain foundational in high-security contexts, they can introduce significant complexity and scalability challenges as mission parameters expand into highly dynamic, multi-domain integrations. 
As defence, emergency response, and critical infrastructure operations increasingly require interoperable, flexible, and cost-efficient communication environments, there is a growing need to understand whether software-based approaches can provide comparable assurance while supporting modern operational requirements.
This SoK characterises a transition to software-based Multi-Domain Data Segregation (MDDS) by integrating systems security, networking, and cryptography. 
Its goal is to consolidate existing research, identify shared architectural patterns, and address the security challenges facing next-generation high-assurance software-based VCS architectures.
By assessing software-defined and virtualized approaches, this SoK supports the development of scalable, high-assurance VCS architectures that facilitate secure real-time coordination across diverse operational domains. Specifically, this SoK examines the security implications of Software-Defined Networking, Network Slicing, Separation Kernels, and Cross-Domain Solutions while addressing challenges posed by quantum computing through Post-Quantum Cryptography (PQC). 
As the first comprehensive synthesis of the shift from hardware-enforced isolation to software-based segregation, this SoK serves as an essential foundation for researchers and industry stakeholders seeking to advance secure, adaptable, and future-ready VCS infrastructures for mission-critical operations.

\keywords{Voice Communications Systems \and Cross Domain Solution \and Software-Defined Networking \and Network Function Virtualization \and Network Slicing \and Separation Kernel \and Air Gap}

\end{abstract}

\section{Introduction}

\begin{figure}[htb!]
\centering
\includegraphics[scale=0.54]{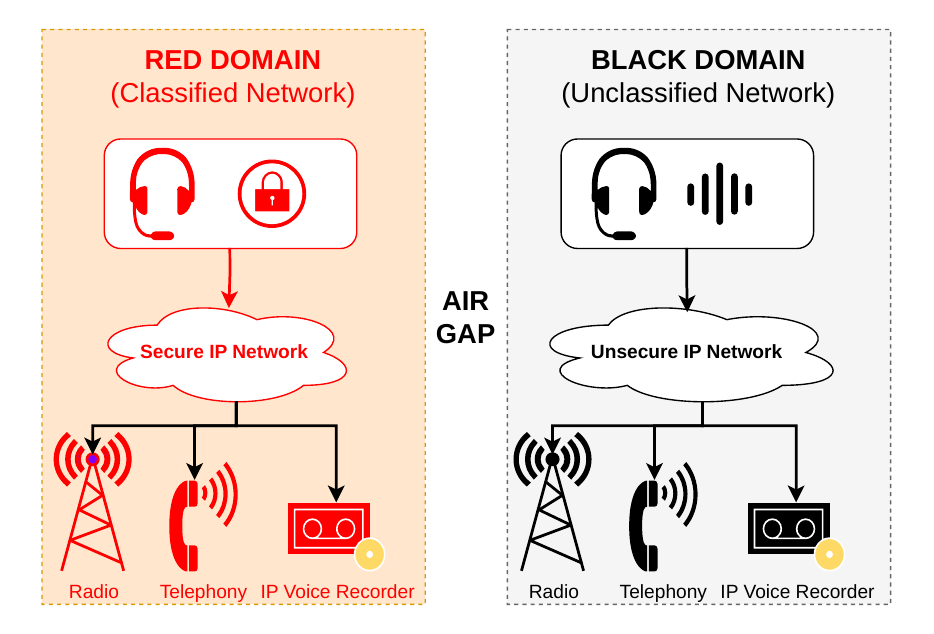}
\caption{Red/Black architecture with strict air-gap isolation, ensuring complete separation between classified and unclassified domains.}
\label{fig:air_gap_architecture}
\end{figure}

Modern operational environments increasingly require communication systems that can support multiple security domains simultaneously \cite{jorquera2022design}. However, traditional air-gapped architectures were built around the assumption of strict physical separation and are typically limited to two domains. While these architectures provide strong security guarantees, their reliance on complex, hardware-only separation makes them costly, inflexible, and difficult to scale when meeting the efficiency demands of contemporary multi-domain operations \cite{na2024study}. As illustrated in Figure \ref{fig:air_gap_architecture}, hardware-enforced isolation relies on completely separate networks to prevent cross-domain leakage of voice data. %Only minimal control data can traverse the Cross Domain Solution (CDS), where it undergoes strict filtering and controlled transfer. 
This reliance on physical duplication often confines systems to just two security domains, revealing a notable limitation when navigating the complexities of modern environments that demand the integration of multiple, diverse security enclaves.

The concept of multiple domains contains distinct security domains, defined as systems operating within specific administrative and security boundaries and governed by a standard security policy \cite{ASD2021CDS}. Within these boundaries, information assets are protected and managed in accordance with established security principles and procedures. This approach segments information environments based on varying trust and security obligations \cite{ASD2021CDS}.

Each security domain is governed by a security policy that outlines information classification, release conditions, and handling requirements \cite{US2005}. For example, NATO uses a five-tier classification system (see Table \ref{tab:nato_classification} for further details). These policies specify who can access domain-specific information and establish necessary security controls to maintain data confidentiality, integrity, and availability \cite{US2005}.

\begin{table}[htb!]
\centering
\caption{NATO security classification levels and their disclosure impact.}
\label{tab:nato_classification}
\small
\begin{tabularx}{\columnwidth}{l X}
\toprule
\textbf{NATO Classification} & \textbf{Description} \\
\midrule
COSMIC TOP SECRET (CTS) 
& Protects information whose disclosure would cause exceptionally grave damage to NATO. \\
\midrule
NATO SECRET (NS) 
& Covers information whose disclosure would cause severe damage to NATO. \\
\midrule
NATO CONFIDENTIAL (NC) 
& Applies to information whose disclosure would damage NATO's interests. \\
\midrule
NATO RESTRICTED (NR) 
& Used for information whose disclosure would be disadvantageous to NATO. \\
\midrule
NATO UNCLASSIFIED (NU) 
& Refers to official NATO information that does not meet classification criteria and may be shared externally when it poses no risk to NATO. \\
\bottomrule
\end{tabularx}
\end{table}

In practice, a security domain can refer to an organisation’s enterprise network operating under a specific classification or to an enclave within a larger network \cite{ASD2021CDS}. External networks, such as the public internet, are separate domains with undefined security controls, posing additional risks, such as malicious ingress, data egress, traffic interception and manipulation \cite{ASD2021CDS,oppliger1997internet,ferguson1998network,ullah2018data}. The concept of security domains is crucial in designing multi-domain environments. The RED/BLACK concept contains the segregation of electrical and electronic circuits, components, equipment, and systems that process classified plain text information, referred to as RED data, from those that manage unclassified information, designated as BLACK data \cite{CNSSI4009}. This distinction is crucial in maintaining the integrity and security of sensitive information by ensuring that classified and unclassified data are kept separate in terms of their electrical signal forms. As illustrated in Figure \ref{fig:air_gap_architecture}, classified information (RED) is kept apart from unclassified information (BLACK) to ensure the distinct management of sensitive data across different systems \cite{C4I}. This segregation is essential for secure communications and data management, especially in military and governmental contexts. As security challenges evolve, multi-domain architectures are increasingly essential to address complex risks.

In this SoK, we examine potential pathways for transitioning from hardware-based isolation to software-based multi-domain data segregation in VCS, with particular awareness to balancing operational scalability and strong security guarantees. Our study is guided by three research questions:

\begin{enumerate}[label=\textbf{RQ\arabic*:}, leftmargin=*]
    \item What software-based methods, frameworks, and architectures have been proposed to enable secure multi-domain data segregation?
    \item What technical and operational limitations have been identified in existing software-based segregation mechanisms?
    \item What research gaps remain, and how can they inform a future roadmap for secure VCS design?
\end{enumerate}

The remainder of this paper is structured as follows. Section \ref{sec:SLR} details the methodology employed for this systematization. Section \ref{sec:reportingSLR} addresses RQ1 by categorizing the 80 reviewed papers into thematic groups and identifies the technical limitations pertinent to RQ2. In addressing RQ3, Section \ref{sec:dis} summarizes the findings through a comparative analysis of the reviewed approaches, highlighting their contributions, limitations, and suitability for secure multi-domain VCS. This section also presents these findings in the form of a layered architectural model for software-based multi-domain systems and assesses the system-level limitations of the current mechanisms. Collectively, these sections highlight a significant research gap: the lack of a unified software solution for operator-centric VCS. Section \ref{sec:con} proposes a roadmap for integrated architectural frameworks and empirical validation to strike a balance between security and real-time performance. Finally, Section \ref{sec:lim} discusses the limitations of this study and outlines directions for future research.

This SoK makes three significant contributions. First, it offers a comprehensive cross-disciplinary synthesis of 80 papers across systems security, networking, cryptography, and critical infrastructure, effectively unifying previously fragmented research on various aspects of VCS security. Second, it presents a structured evaluation framework for approaches to multi-domain data segregation. Finally, it identifies key research gaps and outlines a forward-looking roadmap to facilitate the secure design of next-generation, scalable, software-based multi-domain data segregation for VCS.

\section{Background}

\textbf{Voice Communication Systems} (VCS) function as sophisticated digital switching platforms in high-pressure environments such as air traffic control and military operations \cite{levin2011voice}. They enable operators to communicate with pilots and ground crews by converting voice into digital data transmitted over IP-based networks. Each operator interacts through a dedicated workstation equipped with an user interface and headset, which connects to a distributed call control and routing infrastructure rather than a single central system. To safeguard sensitive information, VCS traditionally employ a Red/Black security architecture, where a strict physical air gap separates classified (Red) and unclassified (Black) domains (see Figure \ref{fig:air_gap_architecture}). This design effectively prevents unintended voice leakage, ensuring that communications remain confidential. Only a minimal amount of control data is permitted to traverse the Cross Domain Solution (CDS), where it undergoes rigorous filtering and controlled transfer. Although this model has traditionally depended on hardware-enforced isolation, current trends are shifting towards software-based and virtualized implementations, which introduce greater flexibility while also presenting evolving security challenges \cite{C4I,daigle2003models,guri2014airhopper}.

\textbf{Cross Domain Solution} (CDS) is a security mechanism that regulates information exchange between networks of different security levels \cite{ASD2021CDS}. By default, it blocks all data flow and only permits communication that complies with established security policies, ensuring confidentiality, integrity, availability, authenticity, and accountability across domains. CDS can be implemented as hardware, software, or both, commonly found in military and high-assurance environments. There are three main types: access solutions (view-only interactions), transfer solutions (controlled data movement), and Multi-Level Security (MLS) solutions (managing users and data across classification levels). Well-designed CDS architectures prevent unauthorized access and cross-domain threats while enabling interoperability \cite{ASD2021,sundaravarathan2024cross,ASD2021CDS}. For detailed information regarding CDS, see Appendix \ref{sec:CDS}.

\textbf{Software-defined networking} (SDN) improves upon traditional networks by separating the control plane from the data plane, allowing centralized management of network intelligence while switches handle traffic forwarding. Its architecture comprises three planes: the Data Plane (forwarding devices), the Control Plane (centralized controller using protocols such as OpenFlow), and the Application Plane (network applications). Through programmability and open standards, SDN minimizes vendor lock-in, streamlines network management, and improves scalability and flexibility \cite{maleh2023,gong2015,singh2017,el2021}. For detailed technical specifications and its architectural details, see Appendix \ref{sec:SDN}.

\textbf{Network slicing} enhances virtualization by creating multiple end-to-end service-specific virtual networks over a shared physical infrastructure. Each slice operates as an independent logical network with tailored control, data, and management properties to meet specific performance, reliability, or security needs. Widely used in 5G and mission-critical systems, slicing supports diverse applications, such as ultra-reliable low-latency services and high-bandwidth media delivery, without requiring separate physical networks. Slices are instantiated from predefined templates and implemented with virtualized network functions on shared hardware. Slicing can be vertical (service-oriented), horizontal (resource partitioning), static (preconfigured), or dynamic (real-time adjustments), allowing for scalable and flexible network management \cite{celdran2019,borsatti2022,YamanySameh2021,iliadis2025qrons}. For comprehensive information on various types of network slicing, refer to Appendix \ref{sec:NetSlicing}.

\textbf{Separation kernels}, first proposed by Rushby in 1981 \cite{rushby1981design}, are minimal, high-assurance microkernels or bare-metal hypervisors that enforce strict isolation between software partitions on shared hardware, minimizing the Trusted Computing Base (TCB). Each partition operates as if on a separate machine, with controlled information flow. This architecture is a cornerstone of the Multiple Independent Levels of Security (MILS) architecture and is widely used in high-assurance domains such as avionics \cite{prisaznuk1992integrated}. Their compact design enables formal verification and certification under standards such as Common Criteria EAL 5 and DO-178B/C, making them essential for safety and security critical systems \cite{alves2006mils,zhao2017high,woodcock2009formal,rtca2011,cc2017,ISO2022}.  For further details on separation kernels and comparisons, please refer to Table \ref{tab:sk_comparison} and Appendix \ref{sec:sepkernels}.

\textbf{Internet Protocol Security} (IPsec) is a network-layer security framework that provides confidentiality, integrity, and authentication for IP communications, primarily used in Virtual Private Networks (VPNs) \cite{tariq2023evaluating}. It operates in two modes: transport mode, which encrypts only the packet payload, and tunnel mode, which encrypts the entire IP packet \cite{iliadis2025qrons,tariq2023evaluating}. To counter quantum threats to traditional cryptography, PQC algorithms such as Dilithium and Kyber can be integrated into IPsec, ensuring quantum-resilient communication while utilizing symmetric encryption \cite{iliadis2025qrons}.

\textbf{Access control} is a vital security mechanism that manages access to physical and digital resources through identification, authentication, authorization, and auditing. It includes physical controls such as locks and biometrics and logical controls such as passwords and Multi-Factor Authentication (MFA). Organizations use various models such as Mandatory Access Control (MAC), Discretionary Access Control (DAC), Role-Based Access Control (RBAC), and Attribute-Based Access Control (ABAC), to meet their security needs. Effective access control is crucial for maintaining confidentiality, integrity, compliance, and protection against data breaches \cite{farhadighalati2025systematic,pci2018}.

Existing multi‑domain data segregation solutions are predominantly hardware‑based, relying on physical separation or duplicated infrastructure to enforce security boundaries. While effective, these approaches are costly, difficult to scale, inflexible to reconfiguration, and poorly suited to modern requirements such as rapid deployment, cloud integration, and AI‑enabled processing. There remains a significant knowledge gap in understanding how software‑based architectures can provide equivalent or higher assurance while improving scalability, maintainability, and adaptability. To address this gap, we undertake a systematic literature review (SLR) to examine existing software‑based frameworks, architectures, methods, and techniques for data segregation, identifying their strengths, limitations, and assurance trade‑offs. This evidence‑driven understanding is essential to inform the design of a robust, high‑assurance software‑defined data segregation architecture that can meet the demands of contemporary multi‑domain operational environments.

\section{Systematic Literature Review}
\label{sec:SLR}

This SLR aims to provide a comprehensive and critical examination of current state-of-the-art methods, frameworks, and architectures designed for secure, software-based multi-domain data segregation. 
The review methodology adheres to the guidelines established by Kitchenham and Charters, which provide a systematic and replicable framework for performing SLRs within the discipline of software engineering \cite{kitchenham2007,del2018}. Following these guidelines, the SLR was structured into three key phases: \textit{planning}, \textit{conducting}, and \textit{reporting}.

\subsection{Planning the Review}
In the planning stage, a structured protocol was developed comprising four key components: \textit{defining research questions}, \textit{constructing search strings}, \textit{identifying data sources}, and \textit{determining study selection criteria}. The primary purpose of this protocol was to minimise researcher bias by clearly outlining these elements prior to the execution of the SLR \cite{kitchenham2007,del2018}.

\subsubsection{Defining Research Questions}
The formulation of research questions (RQs) is a key component of the planning stage, serving as the cornerstone that directs the entire SLR process \cite{kitchenham2007,del2018}. To precisely address our objective of investigating secure multi-domain data segregation, we meticulously aligned our methodology with the established RQs as outlined below:

\textbf{Data Searching:} We utilized the PICOC framework (Population, Intervention, Comparison, Outcome, Context), as outlined in Table~\ref{tab:picoc_elements}, to systematically develop our search strategy. The keywords derived from each PICOC element (detailed in Table~\ref{tab:picoc_keywords}) were carefully selected to create search strings that effectively capture literature on software-based segregation methods (RQ1), their identified limitations (RQ2), and the broader gaps in VCS research (RQ3).

\textbf{Data Extraction:} We developed a targeted data extraction form aimed at capturing evidence that directly addresses our research questions (RQs). Specifically, we extracted structural details of proposed software frameworks and architectures to answer RQ1, documented any explicitly stated technical or operational limitations to address RQ2, and recorded the authors' identified future work and unresolved challenges to inform RQ3.

\textbf{Data Analysis:} Our analysis procedures were designed to clarify the extracted data into clear answers for each research question (RQ). This included categorizing the diverse software-based mechanisms and architectures for RQ1, conducting a thematic analysis of the identified operational bottlenecks and limitations for RQ2, and synthesizing the remaining research gaps to propose a cohesive roadmap for the future design of secure VCS (RQ3).

By employing the PICOC framework to structure our search strings and meticulously aligning our extraction and analysis protocols with Research Questions 1, 2, and 3, we establish a rigorously traceable and transparent SLR process \cite{kitchenham2007,del2018}.

\begin{table}[htb!]
\centering
\caption{PICOC elements, definitions, and SLR applications \cite{kitchenham2007,del2018}.}
\label{tab:picoc_elements}
\small
\begin{tabularx}{\columnwidth}{l p{1.7cm} X}
\toprule
\textbf{Element} & \textbf{Definition} & \textbf{SLR Application} \\
\midrule
Population & Population of interest in the SLR. 
& Software artifacts and systems handling multi-domain environments. \\
\midrule
Intervention & Existing approaches addressing the core problem. 
& Software-based security mechanisms, architectures, design patterns, and cryptographic techniques. \\
\midrule
Comparison & Alternative approaches used for comparison. 
& Baseline security mechanisms (e.g., air-gap networks). \\
\midrule
Outcome(s) & Effects of the interventions on the population. 
& Security, efficiency, and flexibility of data segregation in multi-domain environments. \\
\midrule
Context & Research setting or operational environment. 
& Networked, heterogeneous, mission-critical multi-domain systems supporting real-time voice communication over redundant IP networks. \\

\bottomrule
\end{tabularx}
\end{table}

\begin{table}[htb!]
\centering
\caption{Keyword derivation for research questions and search strings (excluding the Comparison element) \cite{kitchenham2007,del2018}.}
\label{tab:picoc_keywords}
\small
\begin{tabularx}{\columnwidth}{l p{1.7cm} X}
\toprule
\textbf{Element} & \textbf{Primary Phrase(s)} & \textbf{Derived Keyword(s)} \\
\midrule
Population 
& Software artifacts, systems, multi-domain 
& ("multi-domain" OR "multi-enclave" OR "multi-security domain" OR "cross-domain" OR "multi-level") \\
\midrule
Intervention 
& Security mechanisms, architectures, design patterns, cryptographic techniques 
& (security OR secure OR cryptograph* OR guard OR virtualization OR middleware OR TEMPEST OR emanations OR logical) 
AND (mechanism OR architecture OR system OR technique OR method OR approach OR software OR platform OR network OR solution OR segregat* OR separ* OR segmentation OR isolation) \\
\midrule
Outcome(s) 
& Security, efficiency, flexibility, data segregation 
& (confidenti* OR integrity OR efficien* OR flexib* OR "information assurance" OR agility OR reconfigurability OR classification OR "zero trust") \\
\midrule
Context 
& Networked, heterogeneous, mission-critical, voice communication 
& ("command and control" OR "military network" OR "tactical network" OR "battlefield network" OR "coalition network" OR "voice communication" OR "heterogeneous network" OR "voice guard") \\

\bottomrule
\end{tabularx}
\end{table}

\subsubsection{Constructing Search Strings}
\label{subsubsec:searchstring}

The search strings used to identify relevant literature were developed based on the PICOC elements presented in Table \ref{tab:picoc_keywords}. First, Primary Phrase(s) were formulated for each element according to its specific SLR focus (e.g., security mechanisms, architectures, design patterns for the intervention element). Second, these primary phrases were systematically expanded into a comprehensive list of derived keywords by incorporating synonyms, related terms, and common variations (e.g., expanding segregation to separation or partition). Finally, the derived keywords were refined for database searching through the application of Boolean logic (AND / OR) and the use of the asterisk (*) for stemming (e.g., segregat* to capture different word forms). The complete set of search terms and operators is presented under the column Derived Keyword(s) in Table \ref{tab:picoc_keywords}.

The search strategy was refined through an iterative process that drew on the authors' research and industry experience, as well as preliminary search results. Various combinations of keywords and Boolean operators (AND/OR) were tested to assess the relevance and range of the retrieved literature. Throughout this process, certain search expressions yielded excessively broad results, incorporating a substantial number of unrelated studies, while others were overly restrictive and risked omitting pertinent research. To address these challenges, the authors review the preliminary findings and discuss potential enhancements to the search strategy. Based on these discussions, keywords were added, modified, or removed, and Boolean operators were adjusted to strike an appropriate balance between comprehensive search coverage and relevance to the research questions. This iterative refinement process ultimately contributed to the development of the final search string utilized in the systematic literature review.

Following this refinement process, duplicate keywords within each PICOC element's set were removed to ensure consistency. The final search query was then formulated using the logical operators AND and OR, as shown below:

\textbf{Search string:} \textit{
("multi-domain" OR "multi-enclave" OR "multi-security domain" OR "cross-domain" OR "multi-level")
AND
(security OR secure OR cryptograph* OR guard OR virtualization OR middleware OR TEMPEST OR emanations OR logical)
AND
(mechanism OR architecture OR system OR technique OR method OR approach OR software OR platform OR network OR solution OR segregat* OR separ* OR segmentation OR isolation)
AND
(confidenti* OR integrity OR efficien* OR flexib* OR "information assurance" OR agility OR reconfigurability OR classification OR "zero trust")
AND
("command and control" OR "military network" OR "tactical network" OR "battlefield network" OR "coalition network" OR "voice communication" OR "heterogeneous network" OR "voice guard")
}

\subsubsection{Identifying Data Sources}
\label{subsubsec:Identify}

The search was conducted up to November 20, 2025, with no specific starting date, allowing for the inclusion of all relevant studies that contained the specified search terms, regardless of their publication year. The review covered academic databases including IEEE Xplore, SpringerLink, Web of Science, Scopus and the ACM Digital Library. Additionally, Google was used to identify white papers and other publicly available documents relevant to this field. In this SLR, we adopted a structured search strategy, as outlined in Table \ref{tab:SearchStrategy}, to detail the steps undertaken in the study \cite{kitchenham2007,del2018}.

\begin{table}[hbt!]
\centering
\caption{Search Strategy.}
\label{tab:SearchStrategy}
\small
\begin{tabularx}{\columnwidth}{l X}
\toprule
\textbf{Category} & \textbf{Details} \\
\midrule

\multirow{5}{*}{\textbf{Academic Databases}} 
& IEEE Xplore \\
& Springer Link \\
& Scopus \\
& ACM Digital Library \\
& Web of Science \\

\midrule

\textbf{General Databases} 
& Google \\

\midrule

\multirow{5}{*}{\textbf{Target Items}} 
& Journal Papers \\
& Survey Papers \\
& Conference Papers \\
& Book Chapters \\

\midrule

\textbf{Language} 
& English \\

\midrule

\multirow{2}{*}{\textbf{Publication Period}} 
& Start date not set \\
& End date: November 20, 2025 \\

\bottomrule
\end{tabularx}
\end{table}

\subsubsection{Determining Study Selection Criteria}

The search execution phase involved applying the constructed search string across the selected databases to retrieve relevant literature. The resulting papers were then categorised by database, as shown in Table \ref{tab:search_results}.

\begin{table}[htb!]
\centering
\caption{SLR Search Results Categorized by Database \cite{kitchenham2007,del2018}.}
\label{tab:search_results}
\small
\begin{tabularx}{\columnwidth}{X c r}
\toprule
\multicolumn{3}{l}{\textbf{Search String}} \\
\midrule
\multicolumn{3}{p{\dimexpr\columnwidth-2\tabcolsep\relax}}{%
\footnotesize ("multi-domain" OR "multi-enclave" OR "multi-security domain" OR "cross-domain" OR "multi-level") AND 
(security OR secure OR cryptograph* OR guard OR virtualization OR middleware OR TEMPEST OR emanations OR logical) AND 
(mechanism OR architecture OR system OR technique OR method OR approach OR software OR platform OR network OR solution OR segregat* OR separ* OR segmentation OR isolation) AND 
(confidenti* OR integrity OR efficien* OR flexib* OR "information assurance" OR agility OR reconfigurability OR classification OR "zero trust") AND 
("command and control" OR "military network" OR "tactical network" OR "battlefield network" OR "coalition network" OR "voice communication" OR "heterogeneous network" OR "voice guard")
} \\
\midrule
\textbf{Database} & \textbf{Date} & \textbf{Papers} \\
\midrule
IEEE Xplore         & \multirow{5}{*}{20-11-2025} & 48 \\
Springer Link       &                             & 1830 \\
Scopus              &                             & 59 \\
ACM Digital Library  &                             & 771 \\
Web of Science      &                             & 33 \\
\midrule
\textbf{Total}      &                             & \textbf{2741} \\
\bottomrule
\end{tabularx}
\end{table}

During this phase, several key observations were made. First, the number of retrieved studies was considerably large, mainly due to the comprehensive and detailed nature of the search string. Second, because each database search was conducted independently, duplicate records were present among the results. Finally, some papers identified through the search originated from different but potentially relevant contexts, indicating the breadth of coverage achieved by the search strategy. To ensure the relevance and quality of the selected studies, it was therefore necessary to establish and apply strict inclusion and exclusion criteria to refine and narrow down the final set of papers used for analysis, as shown in Table \ref{tab:criteria}.

\begin{table}[hbt!]
\centering
\caption{Overview of Inclusion and Exclusion Criteria Applied During SLR Study Selection \cite{kitchenham2007,del2018}.}
\label{tab:criteria}
\small
\begin{tabularx}{\columnwidth}{p{1.1cm} l X}
\toprule
\textbf{Type} & \textbf{ID} & \textbf{Statement} \\
\midrule

\multirow{7}{*}{Inclusion}
& I1 & Journal papers \\
& I2 & Conference papers \\
& I3 & Book chapters \\
%& I4 & White papers \\
& I4 & Survey papers and systematic literature reviews \\
& I5 & Papers in computer science \\
& I6 & Papers in engineering \\

\midrule

\multirow{4}{*}{Exclusion}
& E1 & Papers not written in English \\
& E2 & Duplicate papers within the search results \\
& E3 & Non-primary research, except relevant survey papers and systematic literature reviews satisfying I4 \\
& E4 & Papers that are not accessible \\

\bottomrule
\end{tabularx}
\end{table}

It is essential to clarify that the inclusion criteria include both primary and secondary research. Primary studies offer original technical contributions and supporting evidence, while survey papers and systematic literature reviews are specifically included under criterion I4 to provide a broader understanding of existing research methodologies, technological advancements, and related approaches to secure multi-domain data segregation. These secondary studies facilitate a more comprehensive and systematic synthesis of the literature by helping to identify research trends, technical limitations, and unresolved challenges. Consequently, exclusion criterion E3 explicitly exempts relevant survey papers and systematic literature reviews that meet the criteria of I4. However, a distinction is preserved between primary and secondary studies during the analysis to ensure that findings synthesized from the existing literature are not misinterpreted as independent experimental evidence. The inclusion criteria encompass various types of publications (I1–I4) and relevant research disciplines (I5–I6), with studies selected based on their relevance to the research questions.

\subsection{Conducting the Review}
\label{sec:conductingSLR}

\subsubsection{Literature Review Execution}

In November 2025, we conducted a systematic search across all digital libraries, including IEEE Xplore, SpringerLink, Web of Science, Scopus, and the ACM Digital Library, as listed in Section \ref{subsubsec:Identify}. These libraries were queried using the search terms defined in Section \ref{subsubsec:searchstring}. A total of 2741 papers were screened using these criteria. Following recommendations from prior studies \cite{booth2021,moher2009}, the criteria were applied iteratively across different reading stages to improve efficiency. The full screening workflow, including the number of papers included and excluded at each phase, is summarised in Table \ref{tab:resultsIn&Ex}. After removing papers not written in English (E1), those that already met the inclusion criteria (I1–I6), and duplicates (E2), the exclusion criteria (E1–E4) were applied to titles (1399 papers), abstracts (884 papers), and the introduction and conclusion sections (316 papers). Full-text screening was then conducted on 247 papers, applying both inclusion and exclusion criteria (I1–I6 and E1–E4) to determine the type of evidence. Ultimately, all 80 papers met the selection criteria and were included in the final dataset. In order to rigorously assess their contributions to secure, software-based multi-domain data segregation, the selected works are systematically organized and elaborated upon across six distinct subsections in Appendix \ref{sec:papers}: Software-Defined Networking (SDN), Network Function Virtualization (NFV), and Network Slicing (Section \ref{subsubsec:SDN and NFV}); Cross-Domain, Multi-Domain, and Heterogeneous environments (Section \ref{subsubsec:CDS}); Command, Control, and Military Systems (Section \ref{subsubsec:C2}); Cybersecurity, Threat Detection, and Policy (Section \ref{subsubsec:policy}); Secure IoT Architecture and Policy Control (Section \ref{subsubsec:IoT}); and Foundational Security and Assurance (Section \ref{subsubsec:general}).

\begin{table}[hbt!]
\centering
\caption{Study selection results after applying inclusion and exclusion criteria \cite{kitchenham2007,del2018}.}
\label{tab:resultsIn&Ex}
\small
\begin{tabularx}{\columnwidth}{X l c c c}
\toprule
\textbf{Step} & \textbf{Criteria} & \textbf{Total} & \textbf{Incl.} & \textbf{Excl.} \\
\midrule

Database search & Search string & 2741 & -- & -- \\
\midrule
Non-English removed & E1 & 2741 & 2721 & 20 \\
\midrule
Apply inclusion criteria & I1--I6 & 2721 & 1486 & 1235 \\
\midrule
Duplicates removed & E2 & 1486 & 1399 & 87 \\
\midrule
Title screening & E1--E4 & 1399 & 884 & 515 \\
\midrule
Abstract screening & E1--E4 & 884 & 316 & 568 \\
\midrule
Introduction-conclusion screening & E1--E4 & 316 & 247 & 69 \\
\midrule
Full-text screening & I1--I6, E1--E4 & 247 & 80 & 167 \\

\bottomrule
\end{tabularx}
\end{table}

\subsubsection{Study Selection and Quality Assessment}

To ensure the reliability and consistency of the study selection process, two reviewers conducted an independent screening of the titles, abstracts, and introduction-conclusion sections, and conducted full-text assessments of potential studies, followed by a comprehensive evaluation of full-text articles based on predefined inclusion and exclusion criteria, as delineated in Table \ref{tab:criteria}. The outcomes of the screening process, along with preliminary findings, were systematically shared with the other authors. Any disagreements regarding study eligibility were addressed collaboratively, taking into account the research objectives, methodological frameworks, technical contributions, and relevance to the overarching research questions. If necessary, candidate studies were re-evaluated to clarify uncertainties, and final decisions on inclusion or exclusion were reached by consensus. This collaborative approach was instrumental in minimizing individual reviewer bias and enhancing the consistency and transparency of the study selection process.

Beyond the criteria for eligibility, the quality of the selected studies was assessed qualitatively based on four principal considerations: (i) relevance to the research questions and the scope of secure multi-domain data segregation; (ii) clarity and transparency in the research methodology employed; (iii) adequacy of supporting evidence, which includes technical analyses, experimental results, or theoretical justifications; and (iv) acknowledgment of the limitations inherent in the research and the unresolved challenges that persist. These considerations established a common framework for evaluating the contributions and limitations of the selected studies, especially given the diverse research methodologies and technical approaches represented in the existing literature.

Moreover, a clear distinction was maintained between primary and secondary research when interpreting the findings. Primary studies were assessed for their original technical contributions, proposed mechanisms, and accompanying analytical or experimental evidence. Conversely, survey papers and systematic literature reviews were primarily evaluated for their synthesis of existing knowledge, identification of research trends, and discussion of technological limitations and remaining research gaps. 
This distinction helped prevent findings reported in secondary studies from being interpreted as independent experimental evidence and supported a balanced assessment of the available literature.

The extracted evidence was subsequently analysed using Braun and Clarke's reflexive thematic analysis \cite{braun2006using} to identify and synthesise recurring themes across the 80 selected studies. Two authors independently reviewed the extracted evidence and generated initial descriptive codes for segments relevant to the research questions. These codes were discussed to explore different interpretations and identify relationships across studies, rather than to establish inter-coder agreement. Candidate themes were then developed, iteratively reviewed, and refined through continued engagement with the underlying evidence. Finally, the themes were defined, named, and mapped to the research questions to synthesise the principal technological approaches, security considerations, and research gaps associated with secure software-based multi-domain data segregation. The resulting thematic structure provides the basis for organising and presenting the review findings in Section~\ref{sec:reportingSLR}.

\subsection{Reporting the Review}
\label{sec:reportingSLR}

To ensure full coverage of all 80 reviewed papers, Table~\ref{tab:sum} maps each paper to its corresponding thematic group and summarises the key gaps identified in the literature, which are further analysed in the following subsections.

\begin{table*}[t]
\centering
\caption{Summary of Reviewed Literature}
\label{tab:sum}
\begin{tabularx}{\textwidth}{p{4.5cm} p{5cm} X}
\toprule
\textbf{Theme} & \textbf{Covered Papers} & \textbf{Identified Gap} \\
\midrule

Software-Defined Networking (SDN), Network Function Virtualization (NFV), and Network Slicing 
& \cite{alfaqawi2023_comprehensive_5g,husen2023_requirements_fin,borsatti2022,you2021_towards6g,maleh2023,keshari2021qos_sdn,tzanakaki2017converged,zhou2016experimental,bonfim2019integrated,wang2018novel,duan2015performance,Bastin2016,Poularakis2021,chowdhary2019sdnsoc,cafini2011security,thottan2019network,ouamri2025,gong2015,singh2017,li2023design,alam2020,YamanySameh2021,zhang2024survey,vilalta2020,celdran2019,zhang2023trusted,de2026cerberus,chang2020,wichary2022network,waller2004policy,muzafar2025sdn,el2021,krishnan2021sdn,chen2016software,berto2011specification,kazmi2023survey,kotulski2018towards,borsatti2022,mahmud2021,gomes2024}
& Lack of support for real-time, operator-centric voice communication; no mechanisms for multi-domain voice mixing, per-domain transmit control, or strict non-interference guarantees. \\

\midrule

Cross-Domain, Multi-Domain, and Heterogeneous 
& \cite{zhang2025crowdrouting,wang2022xauth,steinmetz2012use,khalid2021lightweight,martinez2009access,chen2019games,li2024decision,liu2024pecha,chen2025secret,davoli2016satellite,nichols2024cyber}
& Focus on authentication, routing, and policy enforcement; lack of real-time multi-domain communication, operator interaction, and scalable bidirectional secure voice channels. \\

\midrule

Command, Control, and Military Systems (C2/C4ISR) 
& \cite{Morin2019MethodologyFC,gomes2024,Wal2019UnifiedCI,ahmad2021c3i,domingo2015applied,mishra2017improving,sahu2024military,tuchs2011multiSecurity,ou2021research,rodrigues2024sdn,poltronieri2018secure,saha2023sustainment,wrona2020towards,mahmud2021,rojas2022cloud,kampichler2013}
& Voice communication treated as infrastructure, not operator interface; no support for simultaneous multi-domain voice interaction or secure audio isolation. \\

\midrule

Cybersecurity, Threat Detection, and Policy 
& \cite{li2018multi,goode2007attaining,moore1999construct,zou2016implementation,faragallah2024speech}
& Strong foundations in identity, MLS, and cryptography, but limited to single-domain or application-level systems; no support for multi-domain voice workflows or audio-level separation. \\

\midrule

Secure IoT Architecture and Policy Control 
& \cite{rivera2022enabling,el2021,mao2024controller,jazaeri2021edge,delkhosh2025toward}
& Focus on device- and data-centric security; no support for operator-centric communication, real-time voice mixing, or domain-level audio control. \\

\midrule

Foundational Security and Assurance 
& \cite{esteve2008review,jacobsen2015lightweight,schobel2021test,zhang2024research}
& Provide policy, isolation, and validation principles, but lack application to real-time communication systems, especially multi-domain voice and non-interference guarantees. \\

\bottomrule

\end{tabularx}
\end{table*}

\subsubsection{Software-Defined Networking (SDN), Network Function Virtualization (NFV), and Network Slicing}
\label{subsubsec:SDN}

The list of papers in the Section \ref{subsubsec:SDN and NFV} primarily focuses on the enabling technologies of Software Defined Networking (SDN), Network Virtualization, and Network Slicing as foundational concepts for modern and flexible network management. These papers collectively suggest that software-defined approaches are crucial for moving beyond rigid, traditional air-gapped architectures toward a multi-domain operational environment.

\textit{Theme 1: Foundational SDN Architectures and Principles.}
A significant number of studies delineate the architectural and operational foundations of Software-Defined Networking (SDN), emphasizing centralized control, programmability, and abstraction as essential enablers of contemporary networks. Foundational surveys and architectural analyses \cite{gong2015,singh2017,Bastin2016,wang2018novel,kazmi2023survey} elaborate on SDN layers, controller designs, and application ecosystems. Complementary research investigates controller frameworks, programmability, and optimization strategies \cite{chowdhary2019sdnsoc,Poularakis2021,thottan2019network,li2023design,duan2015performance}, illustrating how SDN facilitates scalable and multi-domain control. Earlier studies on programmable routing and network security \cite{cafini2011security} further underscore the evolution towards software-controlled infrastructures.

Security-focused surveys \cite{maleh2023,chen2016software} assess threats, attack surfaces, and mitigation strategies within SDN, while studies oriented towards Quality of Service (QoS) and performance \cite{keshari2021qos_sdn} demonstrate how SDN can uphold service guarantees in dynamic environments. Collectively, these works establish SDN as a mature paradigm capable of implementing complex policies across multiple domains.

\textit{Key Takeaways:} SDN introduces centralized programmability, fine-grained control, and dynamic policy enforcement, serving as the fundamental enabler for transitioning from rigid, hardware-based isolation to software-defined multi-domain architectures.

\textit{Theme 2: NFV, Virtualization, and Service Function Realization.} 
Network Function Virtualization (NFV) and virtualization enhance Software-Defined Networking (SDN) by decoupling network functions from hardware. This enables flexible deployment and orchestration of services. Comprehensive reviews \cite{bonfim2019integrated,alam2020} and studies focused on specific Virtual Network Functions (VNFs) \cite{zhang2024survey} illustrate how these virtualized functions can be integrated, chained, and managed across distributed infrastructures. These findings emphasize the pivotal role of virtualization in promoting scalability, resource efficiency, and multi-domain isolation.

Additionally, experimental and system-level studies \cite{tzanakaki2017converged,zhou2016experimental} further confirm the viability of merging SDN and NFV within heterogeneous optical and packet networks, reinforcing their practicality in real-world applications.

\textit{Key Takeaways:} NFV and virtualization facilitate flexible service deployment and logical isolation, serving as foundational elements for multi-domain systems that operate over shared physical infrastructure.

\textit{Theme 3: Network Slicing and 5G/6G Infrastructures.}
Network slicing has emerged as a vital mechanism for establishing logically isolated domains tailored to specific service requirements. Foundational and survey works \cite{YamanySameh2021,wichary2022network,kotulski2018towards} articulate slicing architectures, guarantees of isolation, and security considerations. Application-driven studies \cite{celdran2019,borsatti2022,borsatti2022} showcase the implementation of slicing in mission-critical and healthcare scenarios, highlighting its ability to satisfy constraints related to latency, reliability, and security.

Experimental validation \cite{vilalta2020} substantiates the feasibility of resource allocation within slicing environments. Furthermore, broader 5G/6G vision papers \cite{alfaqawi2023_comprehensive_5g,you2021_towards6g,husen2023_requirements_fin} identify slicing as a fundamental capability of next-generation networks.

\textit{Key Takeaways:} Network slicing enables the creation of multiple logically isolated networks using shared infrastructure, effectively creating independent operational environments. This approach supports a variety of applications while ensuring high performance and scalability.

\textit{Theme 4: Security, Policy, and Trust Management.}
Numerous studies concentrate on policy enforcement, trust, and secure management within software-defined environments. Policy-based architectures \cite{berto2011specification,waller2004policy} offer mechanisms for defining and enforcing access and control rules. Security frameworks and trust-based systems \cite{zhang2023trusted,de2026cerberus,muzafar2025sdn,krishnan2021sdn} illustrate how Software-Defined Networking (SDN) can incorporate security policies, Quality of Experience (QoE), and automated management in dynamic settings.

Fine-grained access control mechanisms, such as Hash Flow \cite{chang2020}, exemplify how SDN can implement domain-specific permissions, while IoT-focused security management \cite{el2021} underscores the applicability of SDN in diverse and distributed environments.

\textit{Key Takeaways:} Policy-driven control and security frameworks facilitate programmable enforcement of access, trust, and communication constraints, which are crucial for effectively managing interactions across multiple domains.

\textit{Theme 5: Related Networking Concepts.}
Recent research expands the concepts of software-defined networking (SDN) and slicing to encompass wide-area and multi-domain environments. Surveys on SD-WAN \cite{ouamri2025} and multi-domain control architectures \cite{thottan2019network,li2023design} focus on the management of geographically distributed networks. Studies centered on tactical and military applications \cite{mahmud2021,gomes2024} illustrate how SDN and slicing can facilitate mission-critical, multi-domain communication systems, emphasizing essential requirements such as resilience, interoperability, and secure coordination.

\textit{Key Takeaways:} The application of software-defined networking is increasingly relevant in large-scale, distributed, and mission-critical settings, showcasing its effectiveness for multi-domain operations.

\textit{Gap Analysis:} A consistent conclusion emerges from the reviewed studies. Software-Defined Networking (SDN) Network Functions Virtualization (NFV) and network slicing collectively provide a solid technical foundation for logical multi-domain separation. This enables programmable control, flexible service deployment, and scalable isolation over shared infrastructure. However, the majority of these analyses primarily focus on network-level abstractions, data flows, and service orchestration. Within the defined search scope, none of the identified works address the specific requirements for secure real-time operator-centric VCS. Key capabilities such as simultaneous monitoring across multiple domains, per-domain transmission control, secure voice mixing, and stringent non-interference guarantees have not been considered. Additionally, while slicing and virtualization can approximate logical isolation, they do not deliver the assurance levels necessary to replace traditional air-gapped systems in high-security environments. This highlights a fundamental gap. Despite the potential of software-defined technologies to support multi-domain networking, their application in the realm of secure real-time voice communication remains largely uncharted.

\subsubsection{Cross Domain, Multi-domain, and Heterogeneous}
\label{subsubsec:CD}

Cross domain research investigates techniques and architectures that facilitate controlled interactions between different security domains. This includes mechanisms for authentication, guard-based enforcement, and trust establishment \cite{wang2022xauth,khalid2021lightweight,liu2024pecha,zhang2025crowdrouting}. Collectively, these studies illustrate how secure communication and coordination can be achieved across domain boundaries.

\textit{Theme 1: Cross-Domain Authentication and Trust Establishment.}
A significant body of research focuses on authentication protocols and trust mechanisms within cross-domain environments \cite{wang2022xauth,khalid2021lightweight,liu2024pecha}. These studies propose lightweight, privacy-preserving authentication schemes that leverage a combination of cryptographic primitives (e.g., AES, RSA), zero-knowledge proofs, blockchain integration, and distributed storage solutions. They demonstrate that secure and efficient authentication across domains is achievable, even in highly dynamic environments such as 5G heterogeneous networks and Industrial Internet of Things (IIoT) systems.

\textit{Key Takeaways:}
These works lay a robust foundation for secure, efficient, and privacy-preserving cross-domain authentication. However, it is important to note that they are primarily designed for user or device authentication and do not adequately address real-time, operator-centric communication or stringent domain isolation requirements.

\textit{Theme 2: Cross-Domain Communication and Coordination.}
The exploration of cross-domain interaction is further enhanced through routing and coordination frameworks \cite{zhang2025crowdrouting,davoli2016satellite}. These studies illustrate how trust-aware routing, crowdsourced decision-making, and the integration of satellite and terrestrial networks can facilitate communication across diverse and geographically distributed domains. They underscore the critical importance of scalability, adaptability, and trust management in effectively coordinating resources across multiple domains.

\textit{Key Takeaways:}
These methods promote efficient data exchange and coordination across domains, primarily focusing on network-layer optimization and resource management. However, they fall short of offering mechanisms for secure and simultaneous interaction across multiple domains in real-time communication systems.

\textit{Theme 3: Cross-Domain Policy Enforcement and Guard Architectures.}
Traditional cross-domain solutions depend on guard-based architectures and stringent policy enforcement mechanisms \cite{steinmetz2012use}. These systems facilitate controlled information flow between domains through isolation, inspection, and filtering mechanisms, often bolstered by separation kernels and secure communication protocols such as IPsec.

\textit{Key Takeaways:}
The guard model is unidirectional (classified to unclassified) and designed for constrained management messaging, not for high-throughput, real-time, or bidirectional communication. It assumes static message schemas and predictable traffic, limiting adaptability to dynamic, multi-domain systems. Scalability for simultaneous domains or high-rate streaming applications, such as real-time voice communication, is not considered.

\textit{Theme 4: Multi-Domain Security, Defense, and Resilience.}
Numerous studies focus on broader strategies for resilience and defense in multi-domain systems \cite{chen2019games,li2024decision,nichols2024cyber}. These include game-theoretic models for resilient command and control, zero-trust architectures aimed at preventing lateral movement, and conceptual analyses of the evolution of the cyber domain. Collectively, these works underscore the growing complexity and strategic significance of multi-domain operations.

\textit{Key Takeaways:}
These studies offer valuable insights into resilience, threat modeling, and defense strategies across various domains. However, they tend to be primarily conceptual or network-centric and do not thoroughly address operator-level interactions or the requirements for real-time communication.

\textit{Theme 5: Multi-Domain Access Control and Policy-Based Management.}
The topic of access control and policy enforcement across domains is explored through programmable and agent-based frameworks \cite{martinez2009access}. These approaches illustrate how to systematically manage permissions and interactions across multiple administrative domains.

\textit{Key Takeaways:}
Policy-based management facilitates structured and programmable control of cross-domain interactions. Nevertheless, these mechanisms are not specifically designed for real-time communication control or for ensuring strict isolation between concurrent domain interactions.

\textit{Theme 6: Heterogeneous and Multi-Layer Network Integration.} 
Heterogeneous environments add complexity due to the integration of diverse technologies, protocols, and operational contexts \cite{liu2024pecha,chen2025secret,davoli2016satellite}. Research in this field includes privacy-preserving authentication within heterogeneous networks, adaptive key management for quantum multi-domain systems, and integrated space-air-ground networking. These studies emphasize the need for flexible routing, adaptive cryptographic mechanisms, and coordinated control across multiple layers.

\textit{Key Takeaways:} Heterogeneous networks require robust authentication, flexible routing, and adaptive security policies across diverse technologies. Cryptography, blockchain, and federated monitoring provide essential tools to enable secure cross-domain operations.

\textit{Gap Analysis:} While significant advances have been made in cross-domain authentication, trust, routing, and policy enforcement, none of the studies identified within the defined search scope addresses secure, real-time multi-domain voice communication. The reviewed literature focuses primarily on data exchange, traffic management, resource orchestration, and policy control. None provide mechanisms for operators to simultaneously monitor multiple security domains, nor do they support software-defined domain separation capable of replacing traditional two-domain, air-gapped architectures. Features such as secure voice mixing, per-domain transmit control, and strong non-interference guarantees for multi-domain audio flows remain unaddressed within the examined corpus. Even cross-domain guard identified in this review architectures are limited to low-bandwidth, two-domain environments and cannot meet the latency, throughput, or covert-channel requirements of modern VCS.

\subsubsection{Command, Control, and Military Systems (C2/C4ISR)}
\label{subsubsec:C4ISR}

Research on Command and Control (C2) and C4ISR systems establishes the architectural, operational, and security foundations essential for modern military and mission-critical environments. The studies in this category collectively focus on the design of systems-of-systems, interoperability, secure communications, and the advancement of tactical networks through software-defined, cloud-enabled, and multi-domain paradigms \cite{Morin2019MethodologyFC,gomes2024,Wal2019UnifiedCI,ahmad2021c3i,domingo2015applied,mishra2017improving,sahu2024military,tuchs2011multiSecurity,ou2021research,rodrigues2024sdn,poltronieri2018secure,saha2023sustainment,wrona2020towards,mahmud2021,rojas2022cloud,kampichler2013}.

\textit{Theme 1: Command and Control (C2).}
Foundational research delineates methodologies, architectural frameworks, and interoperability mechanisms for large-scale C2 systems \cite{Morin2019MethodologyFC,gomes2024,Wal2019UnifiedCI}. These studies underscore the importance of mission-oriented analysis, layered system design from hardware to user interface and the integration of diverse components through interoperability adapters and distributed communication models. They frame C2 systems as complex, evolving systems-of-systems that necessitate scalability, modularity, and cross-domain coordination.

\textit{Key Takeaways:}
These works describe how C2 systems can be architected, integrated, and validated as complex, distributed systems-of-systems. They emphasize interoperability, standard data models, and robust networking for collaborative operations. However, communication services, including voice, are treated as underlying infrastructure, without guidance on secure, multi-domain operator voice interfaces.

\textit{Theme 2: Security of C3I/C4ISR and Tactical Networks.} 
A substantial body of research examines the security challenges and defensive measures within military communication systems \cite{ahmad2021c3i,sahu2024military,wrona2020towards,kampichler2013}. These studies analyze vulnerabilities, attack vectors, and countermeasures in C3I/C4ISR environments, advocating for multi-layered and data-centric security strategies. Concepts such as distributed MILS and fine-grained information protection emphasize the significance of strong isolation and controlled information flow across varying classification levels.

\textit{Key Takeaways:}
These studies illustrate that contemporary military systems necessitate robust, multi-layered security and strong isolation assurances. However, they predominantly focus on system-level and data-centric security, often overlooking the real-time communication requirements at the operator level.

\textit{Theme 3: Multi-Domain Security and Information Sharing.} 
Numerous studies have explored secure information exchange and management across various domains \cite{domingo2015applied,tuchs2011multiSecurity,poltronieri2018secure}. These works propose architectures and frameworks designed to enforce release policies, manage multi-domain services, and facilitate controlled information sharing in federated and coalition environments.

\textit{Key Takeaways:}
These approaches illustrate that secure interactions across multiple domains are achievable through effective policy enforcement and management-plane integration. However, the focus remains primarily on data exchange and service coordination, rather than enabling simultaneous, real-time operator interaction across domains.

\textit{Theme 4: SDN, 5G, and Next-Generation Military Networking.}
Emerging research delves into the integration of Software-Defined Networking (SDN), 5G, and cloud/edge technologies within military networks \cite{mishra2017improving,rodrigues2024sdn,mahmud2021,saha2023sustainment,ou2021research,rojas2022cloud}. These studies illustrate how software-defined control, network programmability, and distributed computing can enhance flexibility, scalability, and resilience in tactical environments. They demonstrate that modern communication infrastructures are capable of supporting dynamic mission requirements and multi-domain operations through virtualization and orchestration.

\textit{Key Takeaways:}
These technologies establish a robust foundation for modernizing military communication systems and facilitating flexible, multi-domain operations. However, the focus remains primarily on network services, resource orchestration, and infrastructure-level optimization rather than on communication mechanisms that are operator-facing.

\textit{Gap Analysis:} The reviewed literature in C2/C4ISR architectures and military communications focuses on system-of-systems design, interoperability, distributed networking, and modernized tactical networks using SDN, 5G, and cloud/edge technologies~\cite{Morin2019MethodologyFC,gomes2024,Wal2019UnifiedCI,ahmad2021c3i,domingo2015applied,mishra2017improving,sahu2024military,tuchs2011multiSecurity,ou2021research,rodrigues2024sdn,poltronieri2018secure,saha2023sustainment,wrona2020towards,mahmud2021}. While these works address interoperability, policy enforcement, multi-domain information sharing, and network-level security, they generally treat voice communication as an underlying infrastructure rather than as a critical operator interface. Within our defined search scope, none of the identified studies provide mechanisms for secure, real-time multi-domain voice communication that allow operators to monitor and transmit across multiple classified domains simultaneously. Furthermore, current architectures and security frameworks do not address software-defined domain separation, per-domain transmit control, voice mixing, or strong non-interference guarantees, all of which are essential for modern operator-centric C2 environments. Additionally, research on cloud-native and edge computing infrastructures demonstrates how distributed computing resources, services, and policies can be orchestrated to achieve resilience, high performance, and assurance in mission-critical or military environments~\cite{rojas2022cloud}. This study shows that virtualization and the cloud–edge continuum can enable scalable and flexible multi-domain operations. However, these investigations primarily focus on data flows, resource management, and general service orchestration, neglecting the specific challenges of operator-facing, real-time multi-domain voice communication, including secure voice mixing, per-domain transmit control, and low-latency, high-assurance audio delivery unaddressed within the examined corpus.

\subsubsection{Cybersecurity, Threat Detection, and Policy} 
\label{subsubsec:cyber}

This section synthesizes literature on essential cybersecurity mechanisms, including identity federation, access control, multilevel security architectures, secure communication protocols, and formal assurance frameworks \cite{li2018multi,goode2007attaining,moore1999construct,zou2016implementation,faragallah2024speech}. Collectively, these studies establish foundational principles necessary for enforcing security policies and safeguarding sensitive communications in complex, high-assurance environments.

\textit{Theme 1: Identity Federation and Access Control.}  
The management of identity and authentication across distributed systems is effectively addressed through multi-protocol federation frameworks \cite{li2018multi}. These frameworks facilitate interoperability among various identity standards (e.g., SAML, OpenID) and support unified authentication mechanisms across services, such as Single Sign-On (SSO). They illustrate how consistency in authentication and secure identity propagation can be upheld across heterogeneous environments.

\textit{Key Takeaways:}  
Identity federation offers scalable and interoperable authentication across systems, representing a critical component within secure multi-domain environments. Nevertheless, these mechanisms primarily emphasize user authentication and session establishment and fall short in addressing real-time communication control or domain-level interactions.

\textit{Theme 2: Multilevel Security and Policy Enforcement.}  
The exploration of multilevel security (MLS) and policy-based control is facilitated through middleware-driven architectures and management frameworks \cite{zou2016implementation}. These studies elucidate how centralized control and layered security policies can effectively enforce separation across various classification levels within a unified system.

\textit{Key Takeaways:}  
MLS and policy enforcement mechanisms provide structured control over information flow and access across different security levels. However, their operation is primarily at the system and data levels, without adequately addressing real-time communication flows or operator-centric interactions across multiple domains.

\textit{Theme 3: Secure Communication.}  
Secure communication in mission-critical environments is reinforced by precedence-based communication models and specialized cryptographic techniques \cite{goode2007attaining,faragallah2024speech}. These studies demonstrate how communication priority can be maintained in secure IP networks and how voice data can be safeguarded through advanced encryption methods that integrate time- and frequency-domain transformations.

\textit{Key Takeaways:}  
These approaches ensure confidentiality and the prioritization of critical communications, which are vital for mission-critical systems. However, they primarily focus on securing individual communication channels and do not address the management of multiple simultaneous communications across domains.

\textit{Theme 4: Formal Assurance and Certification.}  
Formal assurance processes are analyzed through methodologies for constructing rigorous security arguments \cite{moore1999construct}. These frameworks facilitate the certification and validation of high-assurance systems by providing structured reasoning about the security properties of systems.

\textit{Key Takeaways:}  
Formal assurance techniques are crucial for validating security-critical systems and achieving certification in high-assurance environments. Yet, they remain largely abstract and do not directly confront the implementation challenges faced in multi-domain communication systems.

\textit{Gap Analysis:}  
A consistent limitation emerges across the reviewed literature \cite{li2018multi,goode2007attaining,moore1999construct,zou2016implementation,faragallah2024speech}. The surveyed studies offer robust foundations for identity management, multilevel security, policy enforcement, secure communication, and system assurance. These mechanisms are effective in enforcing security policies and protecting data within single-domain or application-level contexts. However, within our defined search scope, none of the identified studies address the requirements for secure, real-time multi-domain voice communication. Specifically, capabilities such as simultaneous monitoring of multiple domains, per-domain transmission control, secure voice mixing, and strict non-interference guarantees remain unexamined within this body of work. While identity federation, MLS, and cryptographic techniques provide essential building blocks, they do not encompass operator-centric communication workflows within multi-domain environments. This underscores a fundamental gap between established cybersecurity mechanisms and the requirements for secure, real-time, multi-domain VCS.

\subsubsection{Secure IoT Architecture and Policy Control} 
\label{subsubsec:SecIoT}

This section synthesizes a body of research focused on security, trust, and policy enforcement within Internet of Things (IoT), Industrial IoT (IIoT), and the Internet of Battlefield Things (IoBT) environments \cite{rivera2022enabling,el2021,mao2024controller,jazaeri2021edge,delkhosh2025toward}. Collectively, these studies investigate software-defined networking (SDN)-enabled architectures, edge and fog computing frameworks, trust evaluation mechanisms, and distributed policy enforcement mechanisms across expansive, heterogeneous systems.

\textit{Theme 1: Trust Management.}  
The issue of trustworthiness in distributed IoT and IoBT landscapes is addressed through SDN-integrated trust evaluation frameworks \cite{rivera2022enabling}. These methodologies assess the behavior of devices, enabling the enforcement of trust-aware decisions through centralized control planes. Such approaches facilitate the secure participation of devices in mission-critical scenarios.

\textit{Key Takeaways:}  
The trust evaluation mechanisms examined provide a solid foundation for securing device interactions within large-scale distributed systems. However, the focus is predominantly on device-level assurance and neglects aspects related to operator-level communication or interactions that span multiple domains.

\textit{Theme 2: SDN-Based Security Management and Policy Enforcement.}  
A number of studies have elucidated how SDN facilitates centralized security management and policy enforcement across distributed IoT environments \cite{el2021,mao2024controller}. These architectures effectively decouple control and data planes while integrating access control mechanisms, such as role-based access control (RBAC) and token-based authentication. Policies are enforced through distributed enforcement points, including gateways and edge nodes.

\textit{Key Takeaways:}  
SDN-based control frameworks afford flexible and programmable enforcement of security policies across heterogeneous systems. While these approaches prove effective for access control and overall system management, they fail to address real-time communication flows or multi-domain interactions at the operator level.

\textit{Theme 3: Edge/Fog Computing.}  
The exploration of edge and fog computing paradigms aims to enhance scalability, reduce latency, and bolster resilience in IoT systems \cite{jazaeri2021edge,delkhosh2025toward}. The findings demonstrate that computation and security functions can be situated nearer to devices while maintaining centralized coordination through SDN. Key aspects such as monitoring, scalability, and reliability are underscored in the context of wide-area IoT deployments.

\textit{Key Takeaways:}  
Edge and fog architectures promote scalable and responsive control within geographically distributed environments. Nevertheless, the emphasis remains on data processing and system efficiency, thereby neglecting secure, real-time communication between human operators across domains.

\textit{Gap Analysis:} The reviewed literature in IoT, IIoT, and IoBT security provides strong foundations for trust management, policy enforcement, and distributed control through SDN-based architectures, edge/fog computing, and context-aware security policies~\cite{rivera2022enabling,mao2024controller,jazaeri2021edge,delkhosh2025toward,el2021}. These studies demonstrate how to secure heterogeneous, geographically distributed systems and manage resources and policies across multiple nodes. However, they primarily focus on device- and data-centric security; within our defined search scope, none of the identified studies address operator-facing multi-domain voice communication, including secure real-time voice mixing, per-domain transmit paths, and software-enforced domain separation. Furthermore, within the examined corpus, the concepts of trust evaluation, SDN control, and edge/fog placement have not been applied to the challenges of securing operator-centric multi-domain voice channels.

\subsubsection{Foundational Security and Assurance} 
\label{subsubsec:foundsec}

This cluster addresses high-level mechanisms for policy enforcement, system isolation, and verification/validation in complex networked and operational environments~\cite{esteve2008review,jacobsen2015lightweight,schobel2021test,zhang2024research}. Collectively, these works provide foundational guidance for enforcing resource and access policies, achieving strong isolation, and ensuring system correctness in distributed and multi-domain systems.

\textit{Theme 1: Policy-Based Control.}
Policy-driven methodologies for resource allocation and admission control are analyzed within the context of evolving and next-generation networks \cite{esteve2008review}. These investigations elucidate how access, quality of service (QoS), and system behavior can be governed through dynamic policy frameworks, thereby facilitating adaptable and context-aware control across distributed environments.

\textit{Key Takeaways:}
The implementation of policy-based control offers a flexible mechanism for enforcing access and resource constraints across varied domains. Nevertheless, these approaches predominantly operate at the network and service levels and do not sufficiently address application-specific requirements, such as real-time voice control or domain-specific communication constraints.

\textit{Theme 2: System Isolation and Trusted Computing Foundations.} 
The concept of strong isolation is examined through operating system–level mechanisms, including lightweight capability domains \cite{jacobsen2015lightweight}. These strategies aim to minimize the trusted computing base (TCB) while enforcing fine-grained separation among system components, thus providing a solid foundation for high-assurance systems.

\textit{Key Takeaways:}
Fine-grained isolation mechanisms facilitate robust separation within complex systems, supporting secure multi-domain operations. However, these mechanisms primarily concern process and memory isolation, leaving communication-level separation, especially for real-time data streams, such as voice insufficiently addressed.

\textit{Theme 3: Interoperability and System Validation.}
The verification of system correctness and interoperability is thoroughly investigated through comprehensive testing frameworks and validation methodologies \cite{schobel2021test,zhang2024research}. These studies illustrate how complex systems, including those conforming to military standards, can be validated for compliance, correctness, and operational reliability.

\textit{Key Takeaways:}
Rigorous testing and validation procedures are indispensable for ensuring system reliability and certification in high-assurance environments. Nonetheless, existing methodologies predominantly focus on functional correctness and interoperability, rather than verifying real-time behavior or enforcing non interference within multi-domain communication systems.

\textit{Gap Analysis:} The reviewed literature policy-based control, system isolation, and verification/validation provides foundational methods for enforcing access, resource, and security policies in complex, multi-domain, and military-oriented systems~\cite{esteve2008review,jacobsen2015lightweight,schobel2021test,zhang2024research}. These works demonstrate how high-assurance separation, lightweight capability domains, and rigorous verification and validation methods can support secure, distributed operations. However, the surveyed studies are generic and focus on data or system-level interactions rather than operator-facing voice communication. Specifically, within our defined search scope, no studies were identified that address per-domain audio isolation, one-hot transmit mechanisms, secure voice mixing, or verifiable non-interference for multi-domain voice systems.

\section{Discussion}
\label{sec:dis}

A review of the literature reveals several complementary mechanisms for supporting software-based multi-domain data segregation in real-time VCS. Although Software-Defined Networking (SDN), Network Function Virtualization (NFV), network slicing, separation kernels, and Cross-Domain Solutions provide important capabilities for network separation, platform isolation, and security policy enforcement, their individual contributions address different aspects of the overall system requirements. Within the scope of the reviewed literature, the integration of these mechanisms into a unified, secure, operator-centric multi-domain VCS remains insufficiently evaluated. This section therefore first compares the reviewed approaches to identify their respective strengths and limitations, before examining how their complementary capabilities can be organised into a layered architectural model for software-based multi-domain data segregation.

\subsection{Comparative Analysis of Reviewed Approaches}

The extant literature delineates a range of complementary methodologies aimed at achieving software-based multi-domain data segregation. While these methodologies address various dimensions of security, networking, system isolation, and information management, their suitability for VCS is contingent upon their specific isolation mechanisms, operational functionalities, and respective levels of security assurance. This section aims to elucidate their contributions and limitations through a comparative analysis of the six thematic groups identified in Section \ref{sec:reportingSLR}.

Table~\ref{tab:literature_review} synthesizes the comparative findings, revealing that within our defined search scope, no singular research theme comprehensively addresses all requisite dimensions for secure multi-domain voice communication. The comparison indicates that the reviewed approaches provide complementary building blocks rather than a complete solution for secure multi-domain VCS. Network-level mechanisms facilitate logical separation and traffic management; platform-level mechanisms enforce partitioning and resource isolation; and security services manage authentication, authorisation, and controlled information flow. Operational architectures provide the context for integrating these mechanisms into mission-critical environments. Nevertheless, their combined effectiveness for operator-centric voice communication remains insufficiently established, particularly regarding simultaneous multi-domain monitoring, secure voice mixing, per-domain transmission control, real-time performance, and verifiable non-interference.

\begin{table}[htb!]
\centering
\caption{Summary of Related Research Themes and Limitations for Multi-Domain VCS}
\label{tab:literature_review}
\scriptsize % Smaller font to fit narrow column
\setlength{\tabcolsep}{3pt} % Reduce padding between columns
\begin{tabularx}{\columnwidth}{@{} p{1.6cm} p{1.8cm} p{1.7cm} X @{}}
\toprule
\textbf{\shortstack[l]{Research\\Theme}} & \textbf{\shortstack[l]{Main\\Contributions}} & \textbf{Strengths} & \textbf{\shortstack[l]{Limitations for\\Multi-Domain VCS}} \\
\midrule

SDN, NFV, and Network Slicing (Section \ref{subsubsec:SDN}) & Programmable networking, network virtualisation, logical segmentation, and resource orchestration & Scalability, flexibility, traffic management, and efficient resource utilisation & Logical network isolation alone does not establish audio-level isolation, per-domain microphone control, or end-to-end non-interference \\
\midrule

Cross-Domain, Multi-Domain, and Heterogeneous Systems (Section \ref{subsubsec:CD}) & Authentication, trust management, routing, interoperability, and controlled information exchange & Secure coordination and communication across heterogeneous domains & Limited evaluation of simultaneous real-time voice communication and operator-level audio isolation \\
\midrule

C2/C4ISR Systems (Section \ref{subsubsec:C4ISR}) & Distributed command and control, tactical networking, secure information sharing, and interoperability & Mission-critical operational context, distributed coordination, and resilience & Voice is generally treated as supporting infrastructure rather than an operator-centric service with dedicated multi-domain audio controls \\
\midrule

Cybersecurity, Threat Detection, and Policy (Section \ref{subsubsec:cyber}) & Identity federation, multilevel security, access control, cryptographic protection, and formal assurance & Confidentiality, authentication, policy enforcement, and security management & Limited integration with simultaneous voice routing, audio mixing, and domain-specific transmission control \\
\midrule

Secure IoT Architecture and Policy Control (Section \ref{subsubsec:SecIoT}) & Distributed trust management, SDN-based security policies, and edge/fog computing & Heterogeneous device management, distributed monitoring, and localised enforcement & High latency sensitivity for real-time voice streams and a lack of strict, deterministic operator-level audio separation \\
\midrule

Foundational Security and Assurance (Section \ref{subsubsec:foundsec}) & Trusted computing, system isolation, policy-based resource control, and verification/validation & Reduced TCB, platform-level separation, and foundations for high-assurance systems & Limited integration with application-level voice routing, secure voice mixing, and real-time multi-domain communication \\

\bottomrule
\end{tabularx}
\end{table}

These findings highlight the importance of the layered architecture explored in the subsequent subsections, which investigates how network isolation, platform isolation, and security services can enhance a coordinated multi-domain VCS design. Notably, the proposed integration signifies an architectural approach drawn from existing literature rather than a validated substitute for traditional physical separation. Its efficacy will necessitate further assessment against specific security, performance, and assurance requirements.

\subsection{Network Isolation Layer}
\label{subsec:net}

Network separation technologies play a critical role in VCS, particularly in high-assurance environments where each communication flow must be strictly segregated from the operator position to its designated destination domain. In such contexts, logical and virtualized isolation mechanisms must provide strong guarantees that traffic from one security domain cannot interfere with or leak into another, while still maintaining real-time performance and operational flexibility \cite{el2021}.

\textbf{Software Defined Networking} (SDN) represents a mature and programmable approach to network separation. Large-scale deployments such as Google’s B4 WAN architecture \cite{hong2018b4,jain2013b4} demonstrate that SDN can operate at hyperscale with high reliability and performance. By decoupling the control and data planes, SDN enables centralized policy enforcement, dynamic Access Control List (ACL) updates, and fine-grained micro-segmentation, which are particularly valuable in multi-domain VCS environments. Although centralized control introduces a potential single point of compromise, it also allows rapid security patching and coordinated traffic engineering. From a cost perspective, SDN leverages commodity switching hardware and open-source controllers, making it economically attractive, while its flow-level programmability supports high throughput and responsive traffic management across domains \cite{gong2015,singh2017}. 
Moreover, the reviewed literature underscores the applicability of Software-Defined Networking (SDN) in multi-domain and security-sensitive networking environments. Numerous studies explore SDN architectures, controller frameworks, network security, and multi-domain network management \cite{maleh2023,li2023design,thottan2019network,chowdhary2019sdnsoc}. Notably, research on the performance of SDN controllers and network Quality of Service (QoS) emphasizes the critical nature of effective traffic management and resource allocation in dynamic settings \cite{duan2015performance,bholebawa2018,keshari2021qos_sdn,krishnan2021sdn}. Experimental investigations into software-defined optical and heterogeneous networks further illustrate the viability of programmable network control across various deployment configurations \cite{tzanakaki2017converged,zhou2016experimental}. These capabilities are particularly relevant to multi-domain VCS, where communication flows necessitate dynamic configuration and domain-specific traffic policies. However, despite the promise of SDN for flexible network management and traffic separation, these studies do not independently confirm application-level audio isolation or end-to-end non-interference between concurrent voice channels.

\textbf{Network Virtualization} (NV) further enhances separation by abstracting logical networks over shared physical infrastructure. Widely adopted in cloud platforms such as Amazon Web Services and Microsoft Azure, NV uses overlay encapsulation mechanisms (e.g., VXLAN \cite{mahalingam2014virtual}, GRE \cite{farinacci2000generic}, Geneve \cite{gross2020geneve}) to provide multi-tenant isolation. In VCS deployments, this enables the creation of logically independent communication planes for different operational or security domains while sharing the same hardware. Security in NV depends largely on hypervisor correctness and virtual switch enforcement; although not equivalent to physical air gaps, it has been extensively validated in commercial environments. NV offers high feasibility and scalability with moderate cost, making it a practical solution for segregating traffic in distributed, software-defined architectures \cite{alam2020}. 
Additionally, the literature reviewed explores network virtualization and Network Functions Virtualization (NFV) as complementary strategies for creating flexible, isolated network environments. These approaches enable the independent deployment of network functions and services without reliance on dedicated physical appliances, thus facilitating adaptable service placement, resource allocation, and network management across shared infrastructure \cite{alam2020,bonfim2019integrated}. Such capabilities are particularly pertinent to multi-domain VCS architectures, as they enable the deployment of domain-specific network functions and communication services without requiring separate physical networks for each domain. However, it is important to note that logical separation via virtual networks and virtualized network functions does not inherently ensure end-to-end security isolation. The effectiveness of this method relies on the proper configuration and enforcement of virtual switches, hypervisors, network function instances, and their underlying resource boundaries. Notably, the studies reviewed do not demonstrate that network virtualization alone can assure audio-path isolation or prevent unintended information flows between concurrent voice domains.

\textbf{Network Slicing} (NS), standardized within the 3rd Generation Partnership Project (3GPP) 5G framework, extends virtualization concepts by enabling multiple service-specific logical networks to coexist on a common infrastructure \cite{park2023technology,grings2025nasp}. Each slice can be tailored to distinct performance and reliability requirements, such as ultra-low latency, high bandwidth, or mission-critical reliability, characteristics highly relevant to VCS in multi-domain operations. While slicing enforces logical isolation through virtualization layers, it still relies on shared physical resources, raising assurance considerations in high-security deployments. Nevertheless, it offers efficient resource utilization, strong configurability, and high performance, positioning it as a promising approach for future multi-domain VCS architectures where strict separation and service differentiation must coexist \cite{celdran2019,borsatti2022,YamanySameh2021}.
Additional work on mission-critical network slicing and the ADRENALINE transport-network-slicing testbed investigates differentiated communications support and resource allocation under shared infrastructure \cite{borsatti2022,vilalta2020}. These studies support the feasibility of service-level resource management, but do not establish that a slice alone enforces separation of operator voice input, audio output, and call-control state across security classifications.

\textbf{IPsec} serves as a key mechanism for protecting segregated voice traffic in multi-domain networks by providing network-layer confidentiality, integrity, authentication, and anti-replay protection. Standardized in RFC 4301, IPsec faces long-term security challenges due to quantum computing advancements, particularly from Shor’s algorithm, which threatens widely used asymmetric cryptographic schemes such as RSA and ECDSA \cite{ADS2025crypto,ADS2025ISM}. In response, the development of PQC-IPsec integrates quantum-resistant algorithms into Internet Key Exchange (IKE) processes, driven by the National Institute of Standards and Technology's (NIST) post-quantum standardization efforts \cite{NIST2024crypto}. This includes new Federal Information Processing Standards (FIPS) such as FIPS 203 (ML-KEM) \cite{FIPS2032024crypto} for general encryption and FIPS 204 (ML-DSA) \cite{FIPS2042024crypto} for digital signatures. While PQC-IPsec significantly enhances security for high-assurance VCS, it introduces moderate performance overhead and still requires improvements in implementation and interoperability compared to classical IPsec \cite{iliadis2025qrons}. Nonetheless, it marks a crucial step in securing network architectures in the post-quantum era \cite{tariq2023evaluating}.
When considered collectively, network-layer approaches offer complementary controls rather than equivalent security guaranties. Software-Defined Networking (SDN) and virtualization primarily manage connectivity and resource allocation, while slicing enables service differentiation and IPsec secures data in transit. However, none of these methods independently dictate which security domain is permitted to access captured microphone audio. Therefore, a multi-domain VCS requires the integration of network-, platform-, and application-level enforcement, with demonstrated performance and isolation under concurrent communication workloads.

\subsection{Platform Isolation Layer}
\label{subsec:isolayer}

Platform isolation addresses the boundary between software components that share the same computing hardware. In contrast to network isolation, its enforcement scope includes memory, processors, peripherals, direct memory access (DMA), and inter-partition communication. These properties are relevant to VCS architectures that host services from different security domains on a common device.

\textbf{Separation kernels} provide a high-assurance isolation platform that can replaces traditional air-gapped infrastructures by minimizing the Trusted Computing Base (TCB) and enforcing strict partitioning of system components. Introduced by John Rushby in 1981, this concept underpins the Multiple Independent Levels of Security (MILS) framework and operates as a minimal hypervisor that creates logically isolated partitions on shared hardware \cite{rushby1981design}. Each partition functions autonomously while the kernel controls inter-partition communication, ensuring properties such as data separation, information flow control, temporal separation, and fault isolation, all encapsulated in the NEAT principles (Non-bypassable, Evaluatable, Always-invoked, Tamper-proof). Widely used in high-assurance systems, including avionics, separation kernels (e.g., INTEGRITY-178B and LynxSecure, as shown in Table \ref{tab:sk_comparison}) can support stringent security and safety requirements through appropriate assurance and certification processes Common Criteria (CC) provides a framework for evaluating security properties, whereas DO-178C addresses software development assurance for airborne systems \cite{rtca2011}. Unlike traditional air gaps requiring separate hardware, separation kernels allow multiple secure domains to coexist on the same processor while maintaining strict logical isolation, thus significantly reducing hardware overhead. Their compact size makes them amenable to rigorous mathematical verification, positioning separation kernels as scalable and credible alternatives for secure, multi-domain, real-time communication without sacrificing assurance. 
The reviewed studies provide related but not interchangeable evidence for this platform-level approach. Distributed MILS research investigates partition-oriented designs for advanced air traffic management communication services \cite{kampichler2013}, while lightweight capability-domain research examines decomposition of a conventional operating-system kernel into more isolated components \cite{jacobsen2015lightweight}. These contributions support the architectural rationale for reducing shared trusted components, but they do not amount to a validated multi-domain VCS implementation.

\begin{table}[htb!]
\centering
\caption{Comparison of Separation Kernel Implementations (Sec.: Security, Saf.: Safety, RT: Realtime) \cite{zhao2017high}}
\label{tab:sk_comparison}
\scriptsize % Smaller font to fit narrow column
\setlength{\tabcolsep}{2pt} % Reduce padding between columns
\begin{tabularx}{\columnwidth}{@{} l l c c c X c @{}}
\toprule
\textbf{\#} & \textbf{Name} & \textbf{Sec.} & \textbf{Saf.} & \textbf{RT} & \textbf{\shortstack{Certification/\\Compliance}} & \textbf{\shortstack{Formal\\Methods}} \\
\midrule

\multicolumn{7}{l}{\textbf{Industrial Implementations}} \\
\midrule

1  & PikeOS & \(\checkmark\) & \(\checkmark\) & \(\checkmark\) & DO-178B Level B, IEC 61508 SIL 3, EN 50128 SIL 4, ARINC 653 & \(\checkmark\) \\
2  & VxWorks 653 & \(\times\) & \(\checkmark\) & \(\checkmark\) & DO-178B/C Level A, ARINC 653 & ? \\
3  & VxWorks MILS & \(\checkmark\) & \(\times\) & \(\times\) & SKPP, CC, DO-178C Level A & ? \\
4  & INTEGRITY-178B & \(\checkmark\) & \(\checkmark\) & \(\checkmark\) & DO-178B Level A, CC EAL 6+/SKPP, ARINC 653 & \(\checkmark\) \\
5  & INTEGRITY Mult. & \(\checkmark\) & \(\times\) & \(\times\) & Unknown & ? \\
6  & LynxSecure & \(\checkmark\) & \(\checkmark\) & \(\checkmark\) & CC EAL 7, DO-178B Level A & ? \\
7  & LynxOS-178 & \(\checkmark\) & \(\times\) & \(\checkmark\) & DO-178B Level A, ARINC 653 & ? \\
8  & DDC-I Deos & \(\times\) & \(\checkmark\) & \(\checkmark\) & DO-178B Level A, ARINC 653 & ? \\
9  & AAMP7a & \(\checkmark\) & \(\checkmark\) & N/A & CC EAL 7 & \(\checkmark\) \\
10 & ED & \(\checkmark\) & \(\checkmark\) & \(\times\) & CC & \(\checkmark\) \\
11 & ARLX Hyper. & \(\checkmark\) & \(\times\) & \(\checkmark\) & DO-178B Level A, MILS EAL, IEC 61508 & ? \\

\midrule
\multicolumn{7}{l}{\textbf{Academic Implementations}} \\
\midrule

12 & seL4  & \(\checkmark\) & \(\times\) & \(\times\) & None & \(\checkmark\) \\
13 & OKL4 Micro.  & \(\checkmark\) & \(\checkmark\) & \(\times\) & None & \(\checkmark\) \\
14 & XtratuM  & \(\times\) & \(\checkmark\) & \(\checkmark\) & ARINC 653 & \(\checkmark\) \\
15 & PROSPER & \(\checkmark\) & \(\checkmark\) & \(\checkmark\) & None & \(\times\) \\
16 & Xenon & \(\checkmark\) & \(\times\) & \(\checkmark\) & None & \(\checkmark\) \\
17 & Quest-V & \(\checkmark\) & \(\times\) & \(\checkmark\) & None & \(\times\) \\
18 & Muen  & \(\checkmark\) & \(\checkmark\) & \(\times\) & None & \(\checkmark\) \\
19 & POK  & \(\checkmark\) & \(\times\) & \(\checkmark\) & ARINC 653 & \(\times\) \\
20 & AIR/AIR II & \(\checkmark\) & \(\checkmark\) & \(\checkmark\) & ARINC 653 & \(\times\) \\
\bottomrule

\end{tabularx}
\end{table}

Assurance also requires evidence beyond the existence of partitioning mechanisms. Research on formal arguments for certifiers, interoperability testing against military standards, and validation of operational software models emphasises structured verification and testing \cite{moore1999construct,schobel2021test,zhang2024research}. In a shared-hardware VCS, evaluation should additionally cover device assignment, DMA protection, inter-domain information flow, scheduling interference, shared audio peripherals, and the end-to-end delay of concurrent voice paths. The certification or formal-analysis claims shown in Table~\ref{tab:sk_comparison} concern individual products or research implementations and must be interpreted according to the evaluated product version, configuration, and assurance scope; they do not transfer automatically to an integrated VCS.

\subsection{Security Service Layer}
\label{subsec:seclayer}

\textbf{Access Control} is essential in multi-domain VCS, regulating access to operator consoles, signaling services, and classified voice channels to ensure that only authenticated and authorized personnel, services, or devices can engage in communication across different security domains. It encompasses two dimensions: physical access control, which secures facilities using measures such as biometric authentication, and logical access control, which manages digital resources through techniques such as MFA and access control lists (ACLs). A robust access control framework includes identification to establish unique identities, authentication to verify them, preferably using MFA, authorization to dictate permissible actions based on security policies, and auditing to log activities for traceability. For high-assurance environments, Mandatory Access Control (MAC) is effective, while Role-Based Access Control (RBAC) simplifies privilege management by assigning permissions based on roles, and Attribute-Based Access Control (ABAC) refines access decisions by considering user attributes and environmental conditions. Ultimately, effective VCS access control integrates MAC, RBAC, and ABAC approaches to ensure both security and operational reliability \cite{farhadighalati2025systematic,pci2018}.
The reviewed literature also examines SDN-based access control, authentication and identity federation, and cross-domain trust establishment \cite{chang2020,li2018multi,wang2019identity,wang2022xauth}. Such studies demonstrate approaches to regulating access across distributed services, but do not themselves provide evidence that a permitted operator's microphone signal reaches only the selected voice domain. Policy enforcement must therefore extend beyond call admission to the audio-routing and device-control path.

\textbf{Cross Domain Solutions} (CDS) form the foundation for securely connecting networks at different classification levels in multi-domain VCS. They manage risks associated with information transfer by blocking all inter-domain information flow by default and allowing only explicitly authorized data to pass through security enforcement points, preventing leakage of classified voice streams and information. A secure CDS architecture relies on layered enforcement mechanisms across physical, network, and application layers, guided by key security objectives: confidentiality, integrity, availability, authenticity, and accountability. CDS can bridge Unclassified, Secret, and Top Secret networks for monitoring voice traffic, transferring approved recordings, or consolidating audit data. Without proper CDS controls, interconnections risk exposing sensitive voice data to lower classification levels, increasing unauthorized access potential. CDS technologies include access solutions, transfer solutions, and Multi-Level Security (MLS) systems, supporting secure VCS interoperability. In a software-defined multi-domain VCS architecture, CDS act as a trusted bridge between isolated security domains, enabling operational interoperability while preserving the confidentiality, integrity, and availability of sensitive communications \cite{sundaravarathan2024cross,ASD2021}.
Related studies examine controlled information release between federated military and non-military networks, secure multi-domain information sharing, and data-centric protection in NATO settings \cite{domingo2015applied,poltronieri2018secure,wrona2020towards}. These works help characterise authorised data exchange, but their findings must not be treated as proof of safe mixing or redistribution of live voice streams between classifications.

\textbf{Cross Domain Guard} is essential for enforcing data traffic policies to prevent unauthorized disclosure of classified information by blocking accidental and intentional transmissions with specific indicators \cite{steinmetz2012use}. In a VCS environment, where operational commands and session metadata may contain classification markings, the guard serves as a critical checkpoint for inter-domain exchanges. It scans for \textit{dirty words} such as \textit{secret} and classification tags within signaling data.
When XML formats are used for session control or configuration, the guard validates messages against an approved XML schema, rejecting any that deviate from the specified structure. This schema enforcement is vital for maintaining signaling integrity and operational security in VCS environments \cite{steinmetz2012use}.
Auditability is also crucial. Error messages and rejected transmissions are sent to a dedicated audit component via unidirectional links, ensuring no data flows back into the protected domain. Access to audit records is limited to authorized administrators, maintaining accountability and domain isolation \cite{steinmetz2012use}.
Additionally, to address potential buffer overflow in the guard, an external buffer can be implemented upstream to regulate traffic flow and notify senders of capacity limits, ensuring continuity of operations without compromising the guard’s integrity \cite{steinmetz2012use}. Overall, the Cross Domain Guard facilitates secure interoperability in multi-domain VCS architectures, balancing operational effectiveness with the protection of classified information.

Cryptographic protection and distributed security management provide additional, complementary safeguards. Research on voice encryption and inter-domain key services discusses confidentiality and key provisioning \cite{faragallah2024speech,pekerti2024secure,chen2025secret}, while studies of distributed security management in software-defined and IoT environments examine the coordination of policies across heterogeneous infrastructure \cite{el2021}. These mechanisms contribute to protection of communication channels and networked services, but they do not themselves prevent audio from being copied to the wrong domain, mixed between channels, or transmitted through an unauthorised endpoint.

The approaches reviewed encompass distinct yet interrelated security functions: access control dictates who is permitted to act, separation mechanisms restrict the flow of information, cryptography safeguards data during transmission, and CDS or guards supervise transfers only when cross-domain exchanges are authorized. A secure multi-domain VCS must integrate these functions with clear controls for simultaneous listening, policy-constrained transmission, voice routing, and shared-device management. The available evidence supports a layered design rationale, rather than suggesting that any single reviewed approach has definitively achieved complete end-to-end audio non-interference, security assurance, and real-time performance.

\subsection{Representative Operational Scenario}
\label{subsec:operational_scenario}

To illustrate the potential application of the reviewed technologies, consider a deployable coalition air operations centre supporting multinational military operations at an airfield used by both military and civilian aircraft. An operator may be responsible for coordinating national military activities, exchanging operational information with coalition partners, and communicating with civilian aviation authorities. These interactions may span three distinct security domains: (i) a National Secret Domain containing classified national mission information; (ii) a Coalition Partner Domain supporting authorised communications among allied forces; and (iii) a Civil Coordination Domain handling information approved for communication with civilian air traffic controllers and airport personnel \cite{C4I,JAPCC_2018}.

In this scenario, a software-based multi-domain VCS could enable an appropriately authorised operator to monitor incoming communications from multiple security domains through a single workstation while restricting microphone transmissions to one selected domain at a time. Achieving this capability would require the integration of the network isolation, platform isolation, and security service mechanisms discussed in Sections \ref{subsec:net}, \ref{subsec:isolayer}, and \ref{subsec:seclayer}. Network isolation would support domain-specific communication paths, platform isolation would separate domain-specific processing and resources, and security services would enforce access control, authorised information flows, and domain-specific transmission policies. Such an architecture would need to prevent unauthorised audio mixing and information leakage between domains while maintaining the real-time performance required for mission-critical voice communication \cite{C4I,JAPCC_2018}. Although this scenario is illustrative rather than experimentally validated, it demonstrates how the complementary mechanisms identified in this review could address operational requirements for secure software-based multi-domain data segregation.

\section{Conclusion}
\label{sec:con}

This SoK paper synthesizes the security implications of transitioning VCS from hardware-enforced isolation to a software-based segregation model. It highlights that evolving from traditional air-gapped architectures to software-defined, multi-domain VCS can be achieved through the integration of network separation mechanisms such as Software-Defined Networking (SDN), slicing, network virtualization, as well as separation kernels and cross-domain solutions. By combining PQC with IPsec, the paper underscores the importance of long-term confidentiality, while MILS-compliant kernels provide robust isolation guarantees. Together, these technologies create a high-assurance foundation for secure data segregation, enabling scalable and flexible architectures without compromising mission-critical security or real-time performance. Furthermore, this SoK identifies significant research gaps, notably the scarcity of unified architectural frameworks that systematically integrate these technologies and the limited empirical validation in realistic multi-domain VCS environments. Addressing these gaps will represent the next phase of this research, focusing on designing an integrated framework and developing a prototype to assess security, performance, and deployment trade-offs, thereby bridging the divide between conceptual feasibility and practical implementation.

\section{Limitations and Future Work}
\label{sec:lim}

First, a methodological limitation of this study concerns the scope of the literature search. The search strategy was designed to investigate the broader field of software-based multi-domain data segregation, with selected studies on separation kernels and MILS also examined (see Section~\ref{subsec:isolayer} and Appendix~\ref{sec:sepkernels}) to assess their applicability to secure multi-domain systems. However, because the review was not exclusively focused on MILS, relevant studies employing alternative terminology or domain-specific formulations may fall outside the defined search scope. Future research could build upon these findings by conducting a targeted systematic review of MILS-based architectures to capture additional design patterns applicable to secure multi-domain VCS.

Second, while separation kernels offer a promising mechanism for strong mathematical isolation on shared hardware, their viability as a replacement for traditional air-gapped systems requires deeper theoretical scrutiny. Physical separation and software-based isolation confer fundamentally different security assurances, with the latter depending heavily on the correctness of the separation kernel, underlying hardware, device assignments, and system configuration. Because this review does not establish equivalence between separation-kernel architectures and physical air gaps under all operational conditions, future work should formalize a comprehensive threat model. This model should define relevant adversaries, attack vectors, trust assumptions, and security objectives, explicitly addressing compromised virtual machines, unauthorized information flows, shared memory and Direct Memory Access (DMA) vulnerabilities, side- and covert-channel attacks, misconfigurations, and shared peripheral risks to guarantee that compromising one domain cannot expose another.

Third, as a literature-based investigation, this work does not include an experimental implementation or empirical evaluation of a complete software-based multi-domain VCS. Consequently, the operational feasibility and real-time performance of integrating these technologies remain unverified. Future efforts should therefore develop a representative prototype combining platform-level partitioning, domain-specific network communication, security enforcement, and operator-facing voice services. Experimental testing should validate the accuracy of functional audio routing, ensuring operators can simultaneously monitor authorised incoming audio across multiple domains while transmitting microphone input exclusively to a single selected domain. It should also include real-time performance metrics such as end-to-end voice latency, jitter, packet loss, processing overhead, resource utilisation, and multi-session scalability.

Finally, beyond functional prototyping, this study does not provide the formal security assurance or comparative benchmarking required for high-assurance deployment. Future investigations should focus on generating rigorous security evidence through formal verification of memory, device, and information-flow isolation, complemented by penetration testing, adversarial fault injection, and independent security assessments such as Common Criteria evaluations. Furthermore, direct comparative benchmarking against traditional physically separated architectures is needed to quantify security, latency, and operational trade-offs before software-defined solutions can be adopted as viable alternatives in mission-critical VCS environments.

\begin{acknowledgements}
This initiative partnered C4i Pty Ltd with RMIT University, Melbourne, Australia. 
Authors Peter Vinci, Nick Georghiou, Shane Phillips, and Mathieu Philippe are employees of C4i Pty Ltd, a company that develops secure communication solutions. This study is based on a systematic review of published literature and does not evaluate or endorse any specific commercial product or service. C4i Pty Ltd had no influence on the review methodology, study selection, analysis, interpretation of findings, or conclusions presented in this manuscript.
\end{acknowledgements}

% BibTeX users please use one of
%\bibliographystyle{spbasic}      % basic style, author-year citations
\bibliographystyle{spmpsci}      % mathematics and physical sciences
\bibliography{rmit-c4i}
%\bibliographystyle{spphys}       % APS-like style for physics
%\bibliography{}   % name your BibTeX data base

% Non-BibTeX users please use
%\begin{thebibliography}{}
%
% and use \bibitem to create references. Consult the Instructions
% for authors for reference list style.
%
%\bibitem{RefJ}
% Format for Journal Reference
%Author, Article title, Journal, Volume, page numbers (year)
% Format for books
%\bibitem{RefB}
%Author, Book title, page numbers. Publisher, place (year)
% etc
%\end{thebibliography}

\appendix

\hspace{2cm}
\section{Software Defined Networking (SDN)}
\label{sec:SDN}

Software Defined Networking (SDN) emerged as a necessary response to the longstanding limitations of traditional networks. These legacy systems suffered from the rigidity of vertically integrated devices, where control logic (the decision-making process for directing traffic) and the data forwarding plane (the physical transmission of traffic) were tightly coupled within each router and switch. This inflexibility created challenges, making networks difficult and slow to evolve. Every device required manual configuration, policies were enforced inconsistently, and introducing new protocols often necessitated costly hardware upgrades throughout the entire infrastructure \cite{maleh2023,gong2015,singh2017,el2021}. 

SDN addresses this rigid, device-centric approach by separating the network's control plane from its data plane. Instead of each router making independent routing decisions, SDN consolidates all control logic into a single, logically centralized controller. This change allows network switches to focus solely on forwarding packets according to the rules provided by the controller, which is done instantly and automatically. The result is a programmable network that is easier to manage and much more adaptable to modern demands \cite{maleh2023,gong2015,singh2017,el2021}.

By centralizing control, SDN eliminates the tedious and error-prone task of manually configuring each device. Network policies are set at the controller, which automatically and consistently distributes rules across all switches. This approach removes the need for unique, proprietary management protocols, resulting in a flexible and automated operational model that reduces human error, speeds up deployment, and simplifies configuration and troubleshooting. The SDN controller provides a global network view, allowing it to coordinate essential functions such as access control, flow scheduling, load balancing, and security policies. Achieving this level of consistency is often difficult in traditional distributed networks. Additionally, the abstraction layers of SDN enable applications to view the network as a unified system, enhancing application-driven performance. Advococacy for open standards such as OpenFlow addresses vendor lock-in, allowing easy integration of equipment from multiple vendors. This openness enables researchers and enterprises to test new protocols without hardware modifications, fostering the development of innovative network applications. Such vendor independence has driven significant adoption within modern data centers and cloud networks \cite{maleh2023,gong2015,singh2017,el2021}.

\subsection{SDN Architecture}

The Open Networking Foundation outlines three core ideas that define Software-Defined Networking (SDN) \cite{sasaki2016,sasaki2016control}:

\begin{enumerate}
    \item \textbf{Decoupled Control and Forwarding:} \\
    SDN separates the logic that determines how traffic should be handled from the hardware that actually forwards packets. Even with this separation, the control logic continues to interact with and influence the forwarding devices.

    \item \textbf{Logically Centralized Control:} \\
    Rather than relying on distributed decision-making across individual devices, SDN employs a central controller that maintains a global view of the network. This unified perspective enables consistent policy enforcement and reduces delays caused by local configuration.

    \item \textbf{Programmable Network Abstractions:} \\
    SDN exposes abstracted representations of network resources and state, enabling external applications to program the network without dealing with device-specific details.
\end{enumerate}

These principles are reflected in the SDN architectural design, which is typically organized into three planes:

\textbf{Data plane:}

The Data plane (or Infrastructure layer) is responsible for the high-speed forwarding of network traffic. It comprises the physical network equipment, such as switches and routers, and the internal components and APIs that handle packet processing. The core functionality of a network device in this plane is to receive packets at its ports and execute specific network functions based on rules provided by the Control Plane. This includes actions such as forwarding the packet to an egress port, dropping the packet, or modifying its header \cite{maleh2023,gong2015,singh2017,el2021}.

\textbf{Control plane: }

The control layer of Software-Defined Networking (SDN) is built around the SDN controller, which centralizes network intelligence and operates similarly to a network operating system. The controller manages flow tables on switches and routers through the Southbound API, with OpenFlow being the most widely used protocol. Through a secure channel, the controller installs, updates, and removes forwarding rules on OpenFlow-enabled devices \cite{maleh2023,gong2015,singh2017,el2021}.

A variety of SDN controllers exist, differing in language, threading models, OpenFlow support, and performance characteristics. Key examples include \cite{maleh2023,gong2015,singh2017,el2021}:

\begin{itemize}
    \item \textbf{NOX} \cite{gude2008nox}: The first public SDN controller, implemented in C.
    \item \textbf{POX} \cite{prete2014pox}: A Python-based controller derived from NOX, aimed at improving usability and performance.
    \item \textbf{Maestro} \cite{cai2012maestro}: A multi-threaded controller that distributes tasks efficiently to support parallel processing.
    \item \textbf{Beacon} \cite{erickson2013beacon}: A Java-based, multi-threaded, modular controller developed at Stanford, designed for cross-platform use.
    \item \textbf{SNAC} \cite{controller2011snac}: A controller with a web-based interface for configuring rules and managing events.
    \item \textbf{Floodlight} \cite{bholebawa2018}: A lightweight, high-performance version of Beacon, supported by major vendors such as Intel, Cisco, HP, and IBM.
    \item \textbf{McNettle} \cite{voellmy2012}: Built with Nettle (a DSL in Haskell), designed for low-latency, high-performance operation on multi-core servers.
    %\item \textbf{RISE}: Based on the Trema framework (Ruby/C), supporting large-scale network experimentation and debugging.
    \item \textbf{Ryu} \cite{jayawardena2025}: A Python controller supporting OpenFlow 1.5 and enabling domain separation without VLANs.
    \item \textbf{OpenDaylight} \cite{medved2014}: A Java-based, open-source, cross-platform controller using OSGi and REST APIs, supported by industry partners such as IBM, NEC, Cisco, and Ericsson.
\end{itemize}

\textbf{Application plane: }

The Application Plane is the highest level in the SDN architecture, representing the set of software programs and utilities that configure, control, and monitor the network's behavior. This layer enables the implementation of crucial network management rules and services through various applications, such as Intrusion Detection Systems (IDS), firewalls, and load balancing. Crucially, it promotes automation by interacting with the centralized SDN Controller via northbound APIs, typically using protocols such as REST (Representational State Transfer), allowing service providers to dictate network policy and abstracting the underlying network complexities. Sitting above the Control Plane, the Application Plane uses these APIs to access and interpret network status information provided by the controller, enabling applications to manipulate physical network elements indirectly for sophisticated services such as security monitoring and traffic engineering \cite{maleh2023,gong2015,singh2017,el2021}.

\section{Network Slicing}
\label{sec:NetSlicing}

Network slicing builds on virtualization by creating end-to-end, service-specific virtual networks tailored to different performance or security requirements. Each slice functions as an independent logical network with its own control, data, and management characteristics, even while sharing the same physical infrastructure. This approach is essential in modern systems such as 5G, mission-critical communications, and multi-domain environments. Network slicing enables fine-grained customization and efficient resource allocation, allowing providers to support diverse use cases without deploying separate physical networks \cite{celdran2019,borsatti2022,YamanySameh2021}.

\subsection{Vertical Slicing}

Vertical slicing is a key principle in 5G architecture, allowing for the creation of customized end-to-end logical networks tailored to specific industry needs. Each slice covers multiple network layers, such as access, transport, and core domains, designed for particular use cases such as automotive systems and industrial automation. These slices maintain distinct service-level attributes, including latency, throughput, reliability, and security, ensuring isolation to prevent performance degradation. This isolation is vital for delivering specialized services over shared infrastructure, supporting the Network-as-a-Service (NaaS) paradigm for flexible, Service-Level Agreement (SLA)-driven offerings \cite{iliadis2025qrons}.

\subsection{Horizontal Slicing}

Horizontal slicing involves resource partitioning within a single network layer, such as the radio access network (RAN) or transport network. It enables multiple virtualized functions to operate simultaneously, focusing on resource sharing and management rather than end-to-end service delivery. This approach supports scenarios requiring infrastructure sharing and dynamic scaling, key to Software-Defined Networking (SDN) and Network Functions Virtualization (NFV). For instance, vertical baseband units in the RAN can share hardware while maintaining distinct performance profiles, enhancing network flexibility and cost-effectiveness \cite{iliadis2025qrons}.

\subsection{Static Slicing}

Static slicing configures network slices during the design phase, keeping them fixed throughout operation. Each slice has a predetermined set of resources and performance metrics optimized for specific services. While it simplifies administration, static slicing is less adaptable to demand fluctuations, risking inefficient utilization and making it difficult to accommodate new services without manual reconfiguration. Although important in early network slicing, its limitations are more evident in today's dynamic 5G environments, prompting a shift towards more adaptive solutions \cite{iliadis2025qrons}.

\subsection{Dynamic Slicing}

Dynamic slicing allows networks to create, modify, and remove slices in real-time based on user demand and network conditions. Unlike static slicing, it employs automation and AI-driven management for efficient resource provisioning with minimal human input. This flexibility is crucial for modern applications such as connected vehicles and IoT deployments, enhancing resource efficiency and reducing operational costs. Dynamic slicing requires real-time monitoring and orchestration, representing a significant advancement in network slicing, especially for future 6G networks accommodating variable services and stringent SLAs \cite{iliadis2025qrons}.

\section{Cross Domain Solution (CDS)}
\label{sec:CDS}

Cross Domain Solution (CDS) is a system that includes security mechanisms specifically designed to manage the risks associated with accessing or transferring information between different security domains. 
A CDS is designed to block the flow of information between different security domains by default, permitting only specific data to pass through security enforcement points when it fully complies with the defined security policy. The security-enforcing functions within a CDS can be implemented using separate hardware or software components, and the system’s architecture must ensure that these components cannot be bypassed \cite{ASD2021}.

A secure CDS implementation ensures that the security policies of all connected domains are consistently enforced across every physical and logical layer of the connection. Therefore, it is essential that the information security objectives are clearly understood  before the design or implementation of a CDS begins. These objectives include \cite{ASD2021CDS}:

\begin{itemize}
    \item \textbf{Confidentiality:} Ensuring that information is disclosed only to authorized entities.
    \item \textbf{Integrity:} Ensuring that information and systems cannot be modified without proper authorization.
    \item \textbf{Availability:} Ensuring that systems remain accessible and usable by authorized entities whenever required.
    \item \textbf{Authenticity:} Ensuring that the identity of a user, process, or device is verified before granting access to system resources.
    \item \textbf{Accountability:} Ensuring that the origin and integrity of data can be verified, and that any authenticated actions cannot be repudiated (also known as non-repudiation).
\end{itemize}

The secure design and operation of a CDS can only be achieved if every stage of the system lifecycle produces outputs that are traceable to the security requirements derived from these objectives.

Typical CDS applications include importing publicly available data from an official network into a SECRET-level analysis system, combining multiple classified desktop environments into a single interface, or gathering inputs from various classified systems into a single central auditing platform \cite{ASD2021}.

When CDS controls are not properly enforced, connections between different security domains can be exploited by malicious actors. This can result in unauthorised access, theft or alteration of sensitive information, and the compromise of system or data integrity. Weak enforcement may also enable attackers to bypass critical security functions, disrupt essential systems or services, and use less-protected networks as pathways to reach more sensitive environments \cite{ASD2021}.

The National Cross Domain Strategy and Management Office (NCDSMO) delineates three categories of Cross Domain Solutions (CDS): access, transfer, and Multi-Level Security (MLS) solutions. These classifications are based on the varying types of interactions that CDS devices have with data, which can be categorized according to different security levels or classifications \cite{sundaravarathan2024cross}.

\subsection{Access}

Access refers to a type of Cross-Domain Solution (CDS) used when a client in an untrusted or low-side domain needs to view information from a trusted high-side domain, or vice versa. These solutions allow users to see data without allowing any actual data transfer between the domains, thereby helping to prevent information leaks. For example, an employee at the Arizona State University (ASU) Tempe campus (low side) who needs to view specific data stored at the ASU Research Enterprise (ASURE), which works on defence and space systems (high side), could use an access CDS to securely view that information across domains \cite{sundaravarathan2024cross}.

Before Cross-Domain Solutions (CDS), secure data sharing across domains relied on a manual Swivel Chair Setup. Staff used separate desktops for each domain and retyped information between systems. CDS were created to automate and streamline these labour‑intensive processes for domains with different trust levels (e.g., top secret, secret, unclassified), though their design and accreditation are complex and mostly used in military and commercial settings \cite{sundaravarathan2024cross}.

A CDS can be hardware, software, or both, providing isolation between domains while enabling controlled information flow. An access CDS lets a user in one domain view data in another without transferring it. A transfer CDS securely moves or copies data between domains. An MLS solution labels and manages both users and data across multiple classification or trust levels \cite{sundaravarathan2024cross}.

In everyday language, \textit{accessing} often means both viewing and downloading data, but CDS terminology separates these clearly. With an access CDS, only encrypted keyboard/mouse input goes to the remote desktop and encrypted display data returns; files never leave the remote environment. Thus, in CDS terms, \textit{accessing} can mean either viewing only (access CDS) or requesting/downloading data between domains (transfer CDS), and the distinction is critical for security \cite{sundaravarathan2024cross}.

\subsection{Transfer}

Transfer CDS support secure data exchange between domains by allowing information to be moved or copied across environments with different trust levels. They are generally classified as either unidirectional or bidirectional \cite{sundaravarathan2024cross}.

In a unidirectional solution, data flows in only one direction. A common example is a data diode, which ensures information can travel from machine A to machine B, but never in reverse \cite{sundaravarathan2024cross}.

Bidirectional solutions, on the other hand, allow information to flow both ways, making real-time interaction possible. Guards are an example of this type; they control the exchange of data between machine A and machine B based on predefined rules, enabling interactive communication (such as a Zoom call) while filtering which information is permitted and which is blocked \cite{sundaravarathan2024cross}.

\subsection{Multi-Level Security (MLS)}

Multi-Level Security (MLS) refers to systems that tag both users (subjects) and data (objects) with security labels and apply mandatory security rules, such as those from the Bell-LaPadula model. Using Mandatory Access Control (MAC) and labelling, MLS systems ensure that only appropriately cleared users can access certain information. For example, in an environment with top-secret, secret, and unclassified levels, a user with secret clearance may view secret and unclassified data but cannot access top-secret material, while someone with top-secret clearance can access information at all three levels \cite{sundaravarathan2024cross}.

\subsection{Comparison with Other Security Tools}

Unlike conventional firewalls that filter traffic based on IP addresses, ports, or basic protocol rules, CDS perform deep, content-level inspection of all data transfers. While next-generation firewalls add features such as application awareness and limited content analysis, CDS go much further by examining data at the bit level, validating file structures, detecting hidden content, and enforcing granular rules based on data classification and content type \cite{ASD2021}.

Although data diodes are sometimes compared to CDS, their functionality is limited to one-way data transmission, ensuring information only moves in a single direction to prevent leakage. In contrast, CDS enable secure, bidirectional communication through advanced controls, allowing two-way workflows without compromising strict security boundaries, a key advantage in complex operational environments \cite{ASD2021}.

Virtual Private Networks (VPNs) provide encrypted channels for secure data transit but do not validate content or enforce security policies tied to data sensitivity. Consequently, while VPNs protect data integrity during transmission, they are inadequate for connecting domains with differing security classifications or preventing unauthorized data sharing \cite{ASD2021}.

%Another defining feature of CDS is their adherence to formal certification and accreditation standards. For instance, in the United States, CDS products must be approved by the National Cross Domain Strategy and Management Office (NCDSMO) and comply with rigorous requirements established by the Committee on National Security Systems (CNSS). These evaluations ensure CDS deliver a high level of assurance suitable for safeguarding classified and mission-critical systems \cite{ASD2021}.
Another defining feature of CDS is their adherence to formal certification, assessment, and accreditation requirements, which vary across national security frameworks. In the United States, CDS products are subject to requirements established by the National Cross Domain Strategy and Management Office (NCDSMO), including the evolving Raise the Bar (RTB) initiative, which promotes stronger security measures such as hardware-enforced security and advanced content filtering. These requirements are intended to increase the level of assurance provided by CDS when protecting classified and mission-critical systems. Other nations apply their own standards and regulatory frameworks. For example, in Australia, the Australian Signals Directorate (ASD) provides cross-domain security guidance through the Information Security Manual (ISM), which emphasises appropriate security-enforcing mechanisms, secure architecture and design, and high-assurance components for CDS implementations. \cite{ASD2021}

\section{Separation Kernels}
\label{sec:sepkernels}

The separation kernel concept, introduced by Rushby in 1981 \cite{rushby1981design}, serves to separate kernel verification from the validation of trusted code across distinct components. Its primary goal is to enforce the segmentation of software components while minimizing the Trusted Computing Base (TCB). Security is achieved through both the physical separation of system components and trusted functionalities within these components. This concept laid the foundation for the Multiple Independent Levels of Security/Safety (MILS), as described by Jim et al. in 2006 \cite{alves2006mils}, which emphasizes separation and controlled information flow.

A separation kernel is a small, highly secure, and verifiable type of bare-metal hypervisor or microkernel designed to enforce strict isolation between different software components (partitions) running on the same physical hardware. Its primary function is to create an environment that acts as if each partition is a separate, physically isolated machine, with strictly controlled information flow between them \cite{zhao2017high}.

Initially applied in the avionics sector with Integrated Modular Avionics (IMA) in the 1990s \cite{prisaznuk1992integrated}, separation kernels function as partitioning kernels, focusing primarily on safety concerns. Their limited size allows for formal verification, facilitating thorough correctness assessments. The success of formal methods in both academic research and industrial applications is increasingly evident, as noted by Woodcock et al. in 2009 \cite{woodcock2009formal}. Certified security is typically achieved through Common Criteria (CC) evaluation, mandated by the National Security Agency, which involves formal methods for high-assurance levels, comprehensive security analysis, and formal proofs of model correspondence. The Separation Kernel Protection Profile (SKPP), established by the National Security Agency in 2007, specifically addresses the certification of separation kernels, while safety regulations are governed by RTCA DO-178B \cite{rtca2011} and its 2011 successor, DO-178C \cite{RTCA_DO178C}, which includes a formal methods supplement.

\subsection{Multiple Independent Levels of Security (MILS)}

The separation kernel serves as the foundational layer of a Multiple Independent Levels of Security (MILS) architecture. Its role is to provide a trusted execution environment with minimal, formally verifiable code that enforces strict security policies \cite{zhao2017survey}.

The security requirements for MILS encompass four foundational properties:

\begin{itemize}
    \item \textbf{Data Separation}: Each partition functions as a distinct resource. Applications within one partition are unable to modify the applications or private data of other partitions, nor can they command the private devices or actuators associated with those partitions. This property is also referred to as ``Data Isolation.''
    
    \item \textbf{Information Flow Security}: Information exchanged between partitions must originate from an authenticated source and be directed to authenticated recipients. The source must be verifiable to the recipients. This is known as ``Control of Information Flow.''
    
    \item \textbf{Temporal Separation}: This property enables different components to utilize the same physical resource at different time intervals. A resource is allocated to one component for a designated period, thoroughly cleared, and then reallocated to another component. The services accessed from shared resources by applications in one partition cannot be impacted by those in other partitions. This concept is also known as ``Periods Processing.''
    
    \item \textbf{Fault Isolation}: This property limits damage by ensuring that a failure in one partition does not propagate to other partitions.
\end{itemize}

NEAT refers to the well-known properties associated with separation kernels. The acronym NEAT stands for Non-bypassable, Evaluatable, Always-invoked, and Tamper-proof:

\begin{itemize}
    \item \textbf{Non-bypassable}: Security functions must be inviolable, meaning that components cannot utilize alternative communication paths, including lower-level mechanisms, to circumvent the security monitor.
    
    \item \textbf{Evaluatable}: Security functions should be sufficiently small and simple to allow for rigorous correctness proofs through mathematical verification. This requires components to be modular, well-designed, well-specified, well-implemented, small, and of low complexity.
    
    \item \textbf{Always-invoked}: Security functions must always be operational. This means each access or message is subjected to scrutiny by relevant security monitors, ensuring that checks are conducted not only at the initial access but also for all subsequent interactions.
    
    \item \textbf{Tamper-proof}: The system must control modification rights pertaining to the security monitor's code, configuration, and data. This protection prevents unauthorized alterations, whether arising from subversive actions or poorly written code.
\end{itemize}

While these concepts may seem intuitive, formalizing and proving them can be challenging. Separation kernels are typically verified by demonstrating properties related to data separation, temporal separation, information flow security, and fault isolation.

\section{Selected Papers}
\label{sec:papers}

\subsection{Software-Defined Networking (SDN), Network Function Virtualization (NFV), and Network Slicing}
\label{subsubsec:SDN and NFV}

This section consolidates research on programmable network architectures, including SDN controllers, NFV frameworks, virtual network function (VNF) embedding, network slicing, slice security, and emerging 5G/6G infrastructures for mission-critical and next-generation communication systems.

\begin{enumerate}
    \item A Comprehensive Study on 5G: RAN Architecture, Enabling Technologies, Challenges, and Deployment \cite{alfaqawi2023_comprehensive_5g}
    \item A Survey on Requirements of Future Intelligent Networks: Solutions and Future Research Directions \cite{husen2023_requirements_fin}
    \item Mission Critical Communications Support With 5G and Network Slicing \cite{borsatti2022}
    \item Towards 6G wireless communication networks: vision, enabling technologies, and new paradigm shifts \cite{you2021_towards6g}

\item A comprehensive survey on SDN security: threats, mitigations, and future directions \cite{maleh2023}
    \item A Systematic Review of Quality of Services (QoS) in Software Defined Networking (SDN) \cite{keshari2021qos_sdn}
    \item Converged Wireless Access/Optical Metro Networks in Support of Cloud and Mobile Cloud Services Deploying SDN Principles \cite{tzanakaki2017converged}
    \item Experimental demonstration of software-defined optical network for heterogeneous packet and optical networks \cite{zhou2016experimental}
    \item Integrated NFV/SDN Architectures: A Systematic Literature Review \cite{bonfim2019integrated}
    \item Novel architectures and security solutions of programmable software-defined networking: a comprehensive survey \cite{wang2018novel}
    \item Performance analysis for a service delivery platform in software defined network \cite{duan2015performance}
    \item Programmable, Controllable Networks \cite{Bastin2016}
    \item SDN Data Plane Optimization \cite{Poularakis2021}
    \item SDNSOC: Object Oriented SDN Framework \cite{chowdhary2019sdnsoc}
    \item Security Issues in Programmable Routers for Future Internet \cite{cafini2011security}
    \item The network OS: Carrier-grade SDN control of multi-domain, multi-layer networks \cite{thottan2019network}
    \item A comprehensive survey on software-defined wide area network (SD-WAN): principles, opportunities and future challenges \cite{ouamri2025}
    \item A survey on software defined networking and its applications \cite{gong2015}
    \item A Survey on Software Defined Networking: Architecture for Next Generation Network \cite{singh2017}
    \item Design of General SDN Controller System Framework for Multi-domain Heterogeneous Networks \cite{li2023design}
    \item A Survey of Network Virtualization Techniques for Internet of Things Using SDN and NFV \cite{alam2020}

    \item Network Slicing and Management \cite{YamanySameh2021}
    \item A survey of VNF forwarding graph embedding in B5G/6G networks \cite{zhang2024survey}
    \item Experimental validation of resource allocation in transport network slicing using the ADRENALINE testbed \cite{vilalta2020}
    \item Dynamic network slicing management of multimedia scenarios for future remote healthcare \cite{celdran2019}

    \item A Trusted Remote Data Trading Scheme in Hybrid SDN for Intelligent Internet of Things \cite{zhang2023trusted}
    \item CERBERUS: Towards Secure and Automated Multi-operator Management in B5G Through a Dynamic Policy-Based ZSM Framework \cite{de2026cerberus}
    \item Hash Flow: An Access Control Mechanism for Software Defined Network \cite{chang2020}
    \item Network Slicing Security Controls and Assurance for Verticals \cite{wichary2022network}
    \item Policy based network management in high assurance environments \cite{waller2004policy}
    \item SDN Based Network Management and Security in 6G Networks \cite{muzafar2025sdn}
    \item SDN-based security management of multiple WoT Smart Spaces \cite{el2021}
    \item SDN Enabled QoE and Security Framework for Multimedia Applications in 5G Networks \cite{krishnan2021sdn}
    \item Software-Defined Mobile Networks Security \cite{chen2016software}
    \item Specification of a Policy Based Network Management architecture \cite{berto2011specification}
    \item Survey on Joint Paradigm of 5G and SDN Emerging Mobile Technologies: Architecture, Security, Challenges and Research Directions \cite{kazmi2023survey}
    \item Towards constructive approach to end-to-end slice isolation in 5G networks \cite{kotulski2018towards}
    \item Mission critical communications support with 5G and network slicing \cite{borsatti2022}
    \item Software-Defined Multi-domain Tactical Networks: Foundations and Future Directions \cite{mahmud2021}
    \item Surveying emerging network approaches for military command and control systems \cite{gomes2024}
\end{enumerate}

\subsection{Cross-Domain, Multi-Domain, and Heterogeneous}
\label{subsubsec:CDS}

This section consolidates research addressing interoperability and coordination across separate administrative and security domains (cross-domain), joint operations spanning multiple operational or network domains (multi-domain), and integration across diverse technologies and platforms (heterogeneous environments). These works collectively explore routing, authentication, access control, resilience, guard mechanisms, and inter-domain key management in complex distributed systems.

\begin{enumerate}
\item CrowdRouting: Trustworthy and customized cross-domain routing based on crowdsourcing \cite{zhang2025crowdrouting}
\item XAuth: Secure and Privacy-Preserving Cross-Domain Handover Authentication for 5G HetNets \cite{wang2022xauth}
\item Use of cross domain guards for CoNSIS network management \cite{steinmetz2012use}
\item A lightweight and secure online cross-domain authentication scheme for VANET systems in Industrial IoT \cite{khalid2021lightweight}

\item An Access Control Scheme for Multi-agent Systems over Multi-Domain Environments \cite{martinez2009access}
\item A games-in-games approach to mosaic command and control design of dynamic network-of-networks for secure and resilient multi-domain operations \cite{chen2019games}
\item Decision-Dominant Strategic Defense Against Lateral Movement for 5G Zero-Trust Multi-Domain Networks \cite{li2024decision}

\item PECHA: Privacy-Preserving and Efficient Cross-Domain Handover Authentication for Heterogeneous Networks \cite{liu2024pecha}
\item Secret key rate-adaptive inter-domain key service provisioning in heterogeneous protocol-based multi-domain quantum networks \cite{chen2025secret}

\item Satellite Networking in the Context of Green, Flexible and Programmable Networks \cite{davoli2016satellite}
\item Cyber Progression of the Domains of Warfare \cite{nichols2024cyber}
\end{enumerate}

\subsection{Command, Control, and Military Systems}
\label{subsubsec:C2}

This section consolidates research on Command and Control (C2) and Command, Control, Communications, Computers, Intelligence, Surveillance, and Reconnaissance (C4ISR) systems, focusing on architectures, interoperability, secure communications, cyber defense, and modernized military network infrastructures. These works collectively address system-of-systems design, tactical networking, multi-domain security, and cloud/edge-enabled military operations that support mission-critical decision-making and coordinated defense activities.

\begin{enumerate}
    \item Methodology for Collecting and Analyzing User Requirements: Mission-Oriented Analysis \cite{Morin2019MethodologyFC}
    \item Surveying Emerging Network Approaches for Military Command and Control Systems \cite{gomes2024}
    \item Unified CAMELOT Interoperability Adapters for Existing Unmanned Command and Control Systems \cite{Wal2019UnifiedCI}

    \item A Review on C3I Systems' Security: Vulnerabilities, Attacks, and Countermeasures \cite{ahmad2021c3i}
    \item An applied model for secure information release between federated military and non-military networks \cite{domingo2015applied}
    \item Improving Security in Coalition Tactical Environments Using an SDN Approach \cite{mishra2017improving}
    \item Military Computing Security: Insights and Implications \cite{sahu2024military}
    \item Multi-security domain management integration architecture for end-to-end service management in military networks \cite{tuchs2011multiSecurity}
    \item Research on the Middle Platform Service System of Battlefield Data Governance Information based on 5G Technology \cite{ou2021research}
    \item SDN Supported Network State Aware Command and Control Application Framework \cite{rodrigues2024sdn}
    \item Secure Multi-Domain Information Sharing in Tactical Networks \cite{poltronieri2018secure}
    \item Sustainment of Military Operations by 5G and Cloud / Edge Technologies \cite{saha2023sustainment}
    \item Towards Data-Centric Security for NATO Operations \cite{wrona2020towards}
    \item Software-Defined Multi-domain Tactical Networks: Foundations and Future Directions \cite{mahmud2021}
    \item The Cloud Continuum for Military Deployable Networks: Challenges and Opportunities \cite{rojas2022cloud}
    \item Distributed MILS: A novel approach to advanced ATM communication services \cite{kampichler2013}
\end{enumerate}

\subsection{Cybersecurity, Threat Detection, and Policy}
\label{subsubsec:policy}

This section examines foundational and advanced topics in cybersecurity, highlighting mechanisms for secure communication, identity and access management, multilevel security architectures, and the formal assurance processes required to certify high-assurance systems. It also includes specialised cryptographic techniques and frameworks that support robust security policy enforcement across diverse operational environments.

\begin{enumerate}
    \item A Multi-protocol Authentication Shibboleth Framework and Implementation for Identity Federation \cite{li2018multi}
    \item Attaining Precedence-Based Communications in Secure IP Networks \cite{goode2007attaining}
    \item How to Construct Formal Arguments that Persuade Certifiers \cite{moore1999construct}
    \item Implementation of Multi-level Network Security Management System based Middleware Strategy \cite{zou2016implementation}
    \item Speech cryptography algorithms: utilizing frequency and time domain techniques merging \cite{faragallah2024speech}
\end{enumerate}

\subsection{Secure IoT Architecture and Policy Control}
\label{subsubsec:IoT}

This section brings together research that advances secure communication architectures, robust identity and access management frameworks, and practical methods for enforcing and validating security policies. The selected works explore federated authentication protocols, precedence-based communication in protected IP networks, formal assurance techniques for certifiers, multi-level network security management, secure speech processing, and high-assurance communication models. Collectively, these papers highlight emerging strategies for strengthening trust, control, and resilience across modern distributed systems.

\begin{enumerate}
    \item Enabling Device Trustworthiness for SDN-Enabled Internet-of-Battlefield Things \cite{rivera2022enabling}
    \item SDN-based security management of multiple WoT Smart Spaces \cite{el2021}
    \item A controller-based roadside unit plane architecture for software-defined internet of vehicles \cite{mao2024controller}
    \item Edge computing in SDN-IoT networks: a systematic review of issues, challenges and solutions \cite{jazaeri2021edge}
    \item Toward a Robust WAN-Scale IoT-Fog Networks: A Distributed SDN Security Architecture Encompassing Monitoring, Scalability, Reliability, and Security \cite{delkhosh2025toward}
\end{enumerate}

\subsection{Foundational Security and Assurance}
\label{subsubsec:general}

This section covers general frameworks, security models, and system assurance methods applicable across networks, software, and operational domains.

\begin{enumerate}
    \item A Review of Policy-Based Resource and Admission Control Functions in Evolving Access and Next Generation Networks \cite{esteve2008review}
    \item Lightweight Capability Domains: Towards Decomposing the Linux Kernel \cite{jacobsen2015lightweight}
    \item How to Test Interoperability of Different Implementations of a Complex Military Standard \cite{schobel2021test}
    \item Research on Verification and Validation of Operational Software Program Models \cite{zhang2024research}
\end{enumerate}

\end{document}